\PassOptionsToPackage{table}{xcolor}
\documentclass[sigconf]{acmart}
\usepackage{cleveref}
\usepackage{multirow}
\usepackage{array}
\usepackage{makecell}
\usepackage{amsbsy}
\usepackage{pifont}
\usepackage{enumitem}
\usepackage{stfloats} 

\newcommand{\infoicon}{%
\kern -0.15em
  \raisebox{-0.5ex}{%
    \makebox[0.8em][c]{
      \ooalign{
        \hfil\resizebox{0.97em}{!}{$\bullet$}\hfil\cr
        \hfil\raisebox{0.49ex}{\textcolor{white}{\textbf{\sffamily\small i}}}\hfil\cr
      }%
    }%
  }%
  \kern 0.1em 
}

\newcommand{\mySubsub}[2][]{%
  \par
  \def\temp{#1}%
  \def\nospace{nospace}%
  \addvspace{0.9\baselineskip plus 0.5pt minus 0.2pt}%
  \noindent{\bfseries\large #2}%
  \par
  \nobreak  
  \ifx\temp\nospace
    \addvspace{0pt}%
  \else
    \addvspace{0.1\baselineskip}%
  \fi
  \noindent
  \ignorespaces 
}

\makeatletter
\newcommand{\myAppSection}[3][]{%
  \par
  \in@{nospacebefore}{#1}%
  \ifin@
    \addvspace{0pt}%
  \else
    \addvspace{3ex \@plus 0.5ex \@minus 0.2ex}%
  \fi
  \refstepcounter{section}%
  \noindent{\normalfont\normalsize\bfseries \thesection \ \ \ #2}%
  \label[appendix]{#3}%
  \par
  \in@{nospaceafter}{#1}%
  \ifin@
    \addvspace{0pt}%
  \else
    \addvspace{1.0ex \@plus 0.2ex}%
  \fi
  \noindent\ignorespaces
}

\newcommand{\myAppSubSection}[3][]{%
  \par
  \in@{nospacebefore}{#1}%
  \ifin@
    \addvspace{0pt}%
  \else
    \addvspace{2.5ex \@plus 0.5ex \@minus 0.2ex}%
  \fi
  \refstepcounter{subsection}%
  \noindent{\normalfont\normalsize\bfseries \thesubsection \ \ \ #2}%
  \label[appendix]{#3}%
  \par
  \in@{nospaceafter}{#1}%
  \ifin@
    \addvspace{0pt}%
  \else
    \addvspace{0.8ex \@plus 0.2ex}%
  \fi
  \noindent\ignorespaces
}
\makeatother

\newcommand{\mylabel}[1]{%
  \noindent\textit{\sffamily #1.\,}%
}

\definecolor{MetaBlue}{RGB}{0, 100, 224}
\definecolor{MyBlue}{RGB}{95, 155, 230}
\definecolor{MyGreen}{RGB}{120, 190, 100}
\definecolor{MyOrange}{RGB}{240, 140, 30}

\definecolor{objesturesLink}{HTML}{9C8A73} 

\copyrightyear{2026}
\acmYear{2026}
\setcopyright{cc}
\setcctype{by}
\acmConference[CHI '26]{Proceedings of the 2026 CHI Conference on Human Factors in Computing Systems}{April 13--17, 2026}{Barcelona, Spain}
\acmBooktitle{Proceedings of the 2026 CHI Conference on Human Factors in Computing Systems (CHI '26), April 13--17, 2026, Barcelona, Spain}
\acmDOI{10.1145/3772318.3791742}
\acmISBN{979-8-4007-2278-3/2026/04}

\newcommand{\oj}{\mbox{Objestures}}

\begin{document}

\title{\oj{}: Everyday Objects Meet Mid-Air Gestures for Expressive Interaction}

\author{Zhuoyue Lyu}
\affiliation{%
 \institution{Department of Engineering\\University of Cambridge}
 \city{Cambridge}
 \country{United Kingdom}}
  \email{zl536@cam.ac.uk}

\author{Per Ola Kristensson}
\affiliation{%
 \institution{Department of Engineering\\University of Cambridge}
 \city{Cambridge}
 \country{United Kingdom}}
  \email{pok21@cam.ac.uk}

\begin{abstract}
Everyday object-based interactions (EOIs) and mid-air gesture interactions (MAIs) have been widely explored, yet prior work on their integration often targets narrow use cases or specific technologies, leaving designers and developers with limited guidance that generalizes across diverse EOIs and MAIs. We introduce \oj{}~(``Obj''~+~``Gestures'')---five interaction types spanning EOIs and MAIs, forming a design space for expressive uni- and bimanual interaction. To evaluate the usefulness of \oj{}, we conducted an exploratory user study ($N=12$) on basic 3D tasks (rotation and scaling), which showed performance comparable to the headset's native freehand manipulation. To understand the user experience, we conducted case studies with the same participants across three applications (Sound, Draw, and Shadow), where participants found the interactions intuitive, engaging, and expressive, and indicated interest in everyday use. We further demonstrate the potential of \oj{} across diverse contexts through 30 examples, and discuss limitations and implications.
\end{abstract}




\begin{teaserfigure}
\centering
\includegraphics[width=\textwidth]{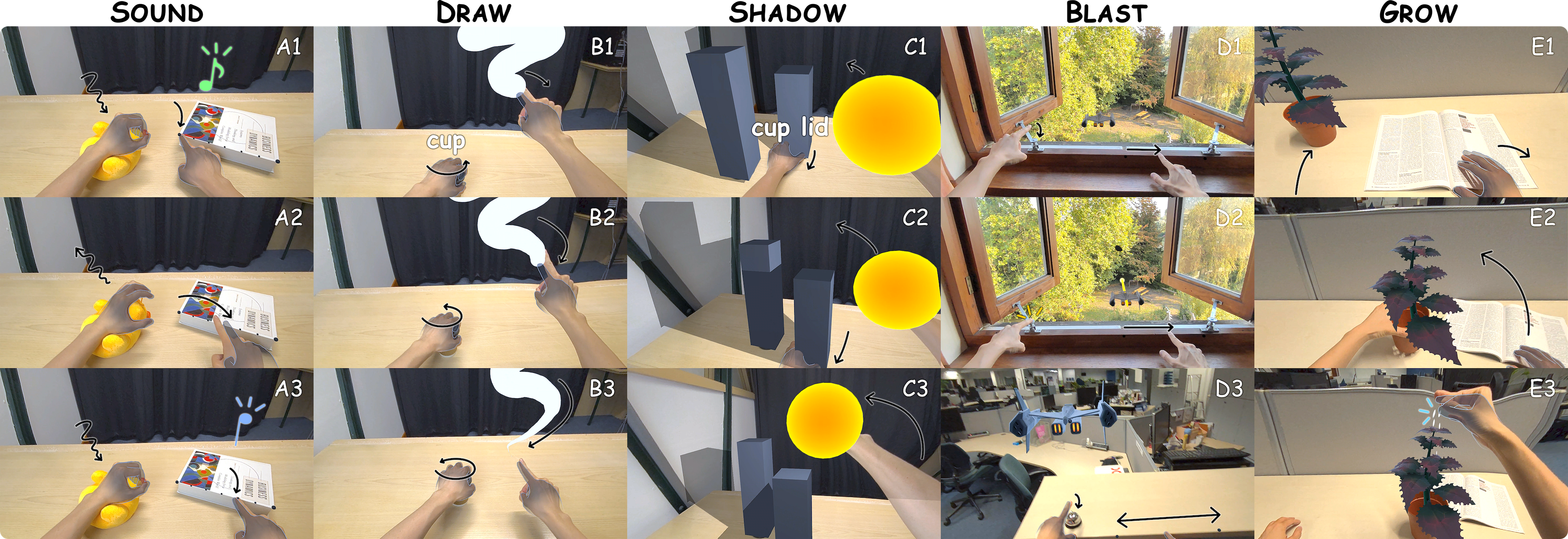}
\vspace{-2em}
\caption{Example applications enabled by \oj{} in mixed reality. \textsc{Sound}: Play a musical instrument by squeezing (A1, A3) and releasing (A2) a stuffed animal to control sound dynamics, while sliding fingers along a book edge (A1--3) to adjust pitch. \textsc{Draw}: Use one hand to draw while rotating a cup with the other to change brush size (B1--3). \textsc{Shadow}: Explore architectural shadows by moving a cup lid that holds a building with one hand (C1--2) while moving the other hand in a fist gesture, overlaid with a virtual sun (C1--3), to observe how the shadow shifts as the sun's position changes. \textsc{Blast}: Play a spaceship shooting game by sliding a finger along a window frame for lateral movement (D1--2), and tapping a window latch knob to fire (D2); this works anywhere, e.g., using a table bell and cabinet edge in an empty office (D3). \textsc{Grow}: Nurture a virtual plant growing in a physical pot by moving the pot with one hand (E1--2) and using a pinching gesture with the other to water it (E3). A--C in Case Studies (\Cref{sec:study2}); D--E in Discussion (\Cref{sec:discussion}).
}
\Description{A grid of 15 first-person perspective captures arranged in five columns and three rows. Each column displays a sequential mixed-reality scenario. Black arrows (both straight and curved) are used across all frames to visually indicate the direction of the user's hand motions. From left to right: in Sound, the user squeezes and releases a stuffed animal to modulate sound dynamics while sliding fingers along a book edge to change pitch. In Draw, one hand draws with the index finger while the other rotates a cup to vary the brush size. In Shadow, inspired by architectural design, the user moves a cup lid that holds a building while the other hand represents the sun, allowing observation of how shadows shift based on the sun's position and direction. In Blast, the user controls a spaceship's lateral movement by sliding a hand along a window frame and fires at obstacles by tapping a window latch knob, with an alternative setup shown using a table bell and cabinet edge in an office. In Grow, the user moves a physical pot that holds a virtual plant with one hand and uses a pinching gesture with the other to water and grow it.}
\label{fig:teaser-full}
\end{teaserfigure}

\maketitle

\section{Introduction}\label{sec:intro} 
With advancements in research~\cite{stellmacher_exploring_2024, kari_reality_2025,lyu_unbounded_2026} and commercial products such as Apple Vision Pro and Meta Quest, mixed reality (MR) is increasingly becoming part of daily life, empowering designers and developers to create experiences that integrate the physical and digital worlds. A crucial aspect of this integration is how to incorporate our living environment into the experience. Studies have recognized that everyday objects can be effectively utilized~\cite{sheng_interface_2006, hettiarachchi_annexing_2016, furumoto_midair_2021} as they inherently possess affordances and offer tangibility and haptic feedback~\cite{gong_affordance-based_2023}. However, they are also shaped by constraints; for example, their functionality depends on the physical availability of objects. Meanwhile, mid-air gestures represent a long-standing human capability~\cite{arendttorp_grab_2023, hosseini_towards_2023}. When integrated into digital experiences, they offer convenience, eliminate the need to carry controllers, and are extensively explored~\cite{villarreal-narvaez_brave_2024, chaffangeon_caillet_glyph_2023, luo_emotion_2024}. However, they are also characterized by constraints, such as the lack of haptic feedback, which can lead to fatigue~\cite{cheema_predicting_2020} and decreased accuracy~\cite{song_hotgestures_2023}.

Everyday object-based interactions (EOIs) and mid-air gesture interactions (MAIs) offer complementary affordances and constraints that open up rich design opportunities when integrated~\cite{gong_affordance-based_2023}. Prior work has explored such integrations, for example, drawing with a pen in one hand while performing mid-air gestures to switch modes~\cite{matulic_pensight_2020}. However, these efforts often target narrow use cases or specific technologies (e.g., pen-based interaction~\cite{aslan_pen_2018}, musical stimulation~\cite{nith_splitbody_2024}, electromyography~\cite{saponas_enabling_2009}). As a result, designers and developers are left with limited guidance that generalizes across diverse objects and gestures, and lack clarity on \emph{how} these modalities can be meaningfully integrated for unimanual and bimanual interaction.

To explore this, we present \oj{}\footnote{\href{https://www.zhuoyuelyu.com/objestures}{{zhuoyuelyu.com/objestures}}} (``Obj'' + ``Gestures''), five interaction types that span both MAIs and EOIs, forming a rich set of playful and expressive interactions (Figures~\ref{fig:designSpace} and~\ref{fig:designSpace-potential}). To evaluate its usefulness, we implemented a prototype and conducted an exploratory user study with 12 participants (\Cref{sec:study1}) to assess whether it can effectively support basic 3D manipulations~\cite{bowman_3d_2004}. We observed performance on object rotation and scaling comparable to the headset's native freehand manipulation, with no significant differences in task completion time, perceived workload (NASA-TLX~\cite{hart_nasa-task_2006}), subjective experience (UEQ-S~\cite{laugwitz_construction_2008}), or preference rankings, and indications of lower error and arm movement. To further understand its user experience, we conducted case studies with the same participants across three example applications---Sound, Draw, and Shadow---that together use all five interaction types. Participants found these applications intuitive, engaging, and expressive, and were interested in using them in daily life. We further illustrate 30 examples to showcase how \oj{} can enrich interactions across diverse contexts, such as everyday tasks, creative arts, education, and gaming (\Cref{fig:designSpace-potential}), and discuss its limitations and implications.

Our contribution therefore is to articulate what has so far remained fragmented across examples~\cite{gong_affordance-based_2023,matulic_pensight_2020,aslan_pen_2018,nith_splitbody_2024,saponas_enabling_2009}. By making these practices comparable and combinable, we enable designers and developers to reason about how and when the complementary properties of EOIs and MAIs can be leveraged. Specifically, we contribute (1) five interaction types with their implementation that form a design space of EOIs and MAIs for expressive interaction; and (2) evaluations and demonstrations of their usefulness, user experience, and potential. Our work is a step towards designing interactions that deepen the connection between the human, the physical world, and the digital experience.

\section{Related Work}\label{sec:related-work}
\subsection{Mid-Air Gesture Interaction}\label{sec:related-work-midair}

From explorations such as \textit{Put-That-There}~\cite{bolt_put-that-there_1980} and \textit{VIDEOPLACE}~\cite{krueger_videoplaceartificial_1985} in the 1980s to current applications in mobile devices~\cite{song_-air_2014, chen_airtouch_2014, zhao_sideswipe_2014}, extended reality (XR)~\cite{shen_fast_2023, pei_hand_2022, lyu_aiive_2021}, public spaces~\cite{walter_strikeapose_2013, ackad_--wild_2015}, and vehicles~\cite{young_designing_2020, shakeri_may_2018, fink_autonomous_2023}, the convenience and intuitiveness of mid-air gestures have been well acknowledged~\cite{song_handle_2012, arora_magicalhands_2019, villarreal-narvaez_brave_2024}. Particularly in XR, they have found expressive uses in areas such as typing~\cite{shen_fast_2023, shen_personalization_2022, markussen_vulture_2014}, sonification~\cite{muller_boomroom_2014, lyu_aiive_2021}, animation~\cite{arora_magicalhands_2019, jiang_handavatar_2023, sayara_gesturecanvas_2023}, presentation~\cite{hall_augmented_2022, liao_realitytalk_2022}, and expressive interactions~\cite{song_hotgestures_2023, pei_hand_2022}.

Throughout this process, a wide range of gestures have been categorized~\cite{hosseini_towards_2023, villarreal-narvaez_brave_2024, aigner_understanding_2012, chaffangeon_caillet_glyph_2023, luo_emotion_2024}, with detailed investigations such as how different user groups (e.g., children~\cite{remizova_exploring_2023}, wheelchair users~\cite{bilius_understanding_2023}, people with visual impairments~\cite{dim_designing_2016, fink_autonomous_2023}, indigenous people~\cite{arendttorp_grab_2023}) and usage postures (e.g., reclining on a sofa~\cite{veras_elbow-anchored_2021} or standing with hands at the sides~\cite{liu_gunslinger_2015}) affect gesture preferences and needs. Research has also investigated mediums for capturing gestures, ranging from on-body devices (e.g., fingertip wearables~\cite{yang_magic_2012, chan_fingerpad_2013, xu_tiptext_2019}, rings~\cite{vatavu_gesturing_2021, vatavu_ifad_2023}, wristbands~\cite{xu_enabling_2022, iravantchi_beamband_2019, devrio_discoband_2022, kim_digits_2012}, forearm bands~\cite{saponas_enabling_2009, saponas_making_2010}) to external devices (e.g., cameras~\cite{liao_realitytalk_2022, song_-air_2014, harrison_-body_2012} and sensors~\cite{min_seeing_2023}).

However, while MAIs offer benefits, constraints such as the lack of haptic feedback can lead to a loss of agency~\cite{cornelio_martinez_agency_2017}, speed~\cite{vuibert_evaluation_2015}, and precision~\cite{song_hotgestures_2023}. Instead of treating these as problems to fix with solutions such as gloves~\cite{gu_dexmo_2016, wang_toward_2019, shen_fluid_2023}, ultrasound~\cite{rakkolainen_survey_2021, shakeri_may_2018}, or muscle stimulation~\cite{lopes_adding_2018, pfeiffer_let_2016}, some see them as design opportunities to extend the interaction space~\cite{matulic_pensight_2020, gong_affordance-based_2023}. We follow this line of inquiry, leveraging EOIs to provide tactile grounding.

\subsection{Everyday Object-Based Interaction}\label{sec:related-work-objects}
From the mouse's invention~\cite{english_display-selection_1967} to the proliferation of various physical input devices~\cite{frohlich_cubic_2000, grossman_interface_2003, saponas_pockettouch_2011, vogel_conte_2011, satriadi_tangible_2022}, the efficiency and user experience benefits of tangible~\cite{ishii_tangible_1997}, graspable~\cite{fitzmaurice_bricks_1995} tools have been well recognized~\cite{fishkin_taxonomy_2004, jacob_reality-based_2008, zhu_phoneinvr_2024}. However, creating specialized hardware remains challenging~\cite{bacher_defsense_2016}, leading to explorations into repurposing existing items. For electronic devices (e.g., keyboards~\cite{schneider_reconviguration_2019}, smartphones~\cite{mohr_trackcap_2019, zhu_touching_2022, matulic_phonetroller_2021, stellmacher_exploring_2024}, tablets~\cite{hubenschmid_stream_2021}, smartwatches~\cite{laput_viband_2016}, soft electronics~\cite{olwal_e-textile_2020, olwal_io_2018}), repurposing often involves leveraging or modifying their intrinsic functionality. For non-electronic everyday objects (e.g., sponges~\cite{sheng_interface_2006}, pens~\cite{corsten_instant_2013}, cups~\cite{seo_gradualreality_2024}, bottles~\cite{hettiarachchi_annexing_2016}, balloons~\cite{furumoto_midair_2021}, metal wires~\cite{li_anicraft_2024}, fruits~\cite{takahira_insitutale_2025}, furniture~\cite{andrei_take_2024, fang_vr_2023, sato_touche_2012, kim_exploration_2022}), it mainly relies on their form factors and affordances. Our work mainly focuses on the latter.

The appeal of everyday objects lies in their ready availability~\cite{henderson_opportunistic_2008} and their capacity to represent users' personal, intimate lives~\cite{ambe_technology_2017}. Given users' diverse preferences for these objects~\cite{greenslade_using_2023} and the varying affordances they offer, research has contributed guidance~\cite{gong_affordance-based_2023} and systems~\cite{hettiarachchi_annexing_2016, jain_ubi-touch_2023} for identifying appropriate objects, and authoring tools for employing them across applications~\cite{corsten_instant_2013, monteiro_teachable_2023, he_ubi_2023}.

A key characteristic of EOIs is their inherent dependence on the object's physical presence. For instance, while \citet{gong_affordance-based_2023} explored twisting a cup to rotate a virtual car model, this interaction breaks if the cup is absent. We mitigate this physical dependency by formalizing the integration of MAIs.

\subsection{Bimanual Interaction}\label{sec:bimanual-interaction}
From the evidence of tool use dating back 3.39 million years~\cite{mcpherron_evidence_2010} to various activities people perform daily using both hands, whether symmetric (e.g., pulling a rope) or asymmetric (e.g., writing)~\cite{bowman_3d_2004}, humans have long relied on the efficiency of bimanual coordination. Pioneering works such as Schlesinger's natural grip taxonomies~\cite{schlesinger_mechanische_1919, mackenzie_grasping_1994}, Buxton and Myers' two-handed input~\cite{buxton_study_1986}, and Guiard's Kinematic Chain Theory~\cite{guiard_asymmetric_1987}, have laid the foundation for a rich body of HCI research~\cite{hinckley_cooperative_1997, wagner_bitouch_2012, smith_dual_2023} that expands upon these principles.

Prior work falls into two groups~\cite{saponas_enabling_2009}: \textit{hands-free} and \textit{hands-busy}. Hands-free interactions utilize mid-air gestures (e.g., pinching~\cite{song_hotgestures_2023, wang_6d_2011}, making a fist~\cite{surale_experimental_2019}, pointing~\cite{jiang_handpainter_2021}) and spatial transformations (e.g., rotation~\cite{song_handle_2012}, translation~\cite{arora_magicalhands_2019, mendes_mid-air_2014}) performed symmetrically or asymmetrically. Hands-busy interactions involve devices (e.g., mice and keyboards~\cite{odell_toolglasses_2004}, wearables~\cite{webb_wearables_2016}, pens~\cite{hinckley_pen_2010, romat_style_2022, zeleznik_hands-math_2010}, touchpads~\cite{adams_sonicexplorer_2014, guimbretiere_bimanual_2012}, smartphones~\cite{kari_handycast_2023, wacker_arpen_2019, tsandilas_interpreting_2012}, tablets~\cite{avery_introducing_2018, hinckley_pen_2010, romat_style_2022}, large displays~\cite{xia_writlarge_2017, webb_wearables_2016, reisman_screen-space_2009}). These interactions can be symmetric, with hands performing identical actions synchronously~\cite{kyto_using_2019} or asynchronously~\cite{kari_handycast_2023} on the same device; or asymmetric, with hands engaging in distinct actions on the same~\cite{wagner_bitouch_2012, kaser_fingerglass_2011, avery_introducing_2018, surale_experimental_2017, hayatpur_plane_2019} or different~\cite{pfeuffer_thumb_2017, wacker_arpen_2019, song_mouselight_2010, hinckley_pen_2010, romat_style_2022} devices.

However, most research has focused on either hands-free or hands-busy; efforts that explore both often focus on specific objects (e.g., pens~\cite{aslan_pen_2018,hinckley_pen_2010,matulic_pensight_2020}) or technologies (e.g., musical sensing~\cite{saponas_enabling_2009} and stimulation~\cite{nith_splitbody_2024}), offering limited generalizable guidance. We address this by formalizing a design space that unifies these hands-free and hands-busy interactions.

\section{Objestures}\label{sec:objestures}
\begin{figure*}
    \centering
    \includegraphics[width=1\linewidth]{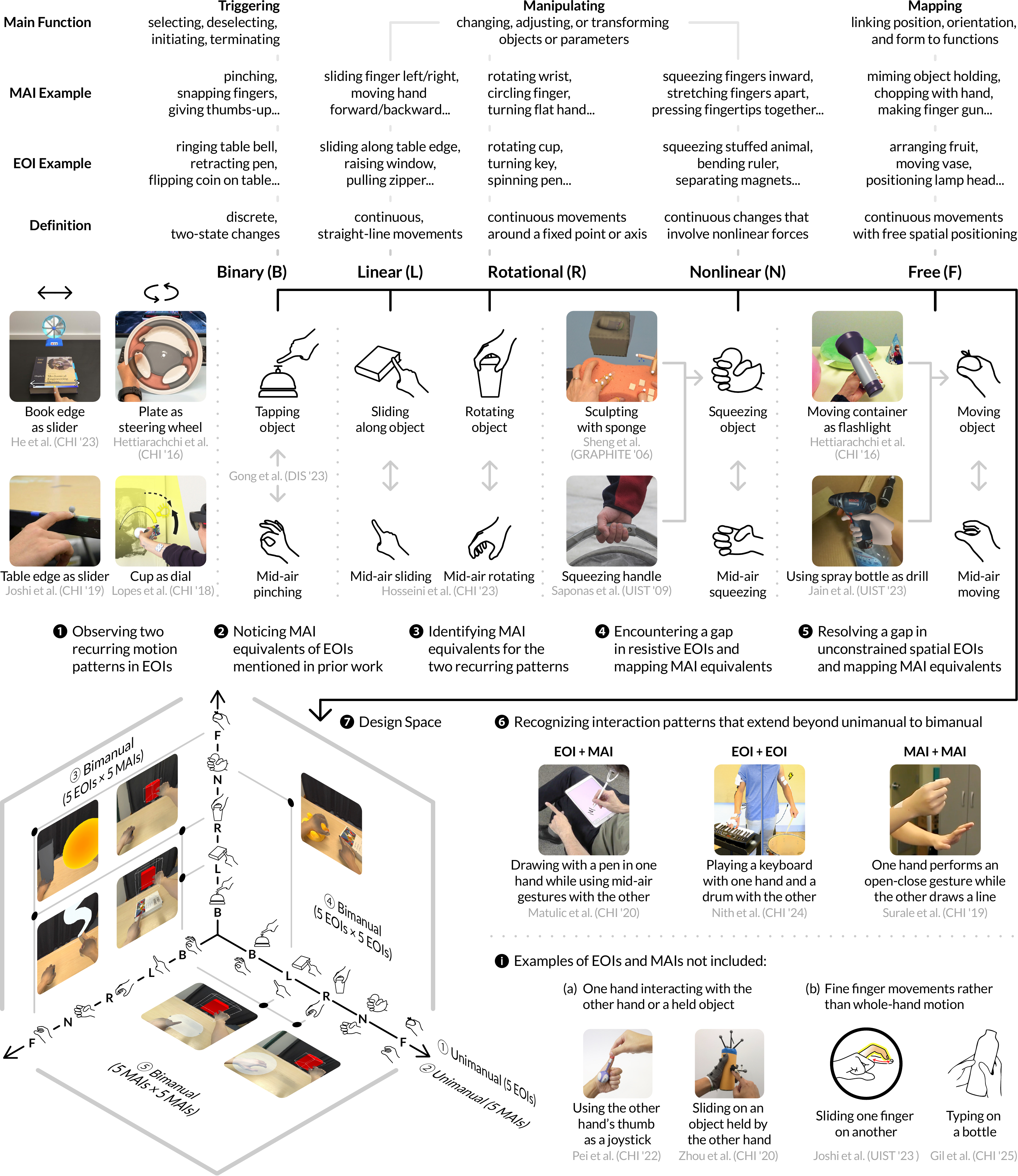}
                \caption{Formation of the five interaction types (Binary, Linear, Rotational, Nonlinear, and Free), as well as the design space. Follow \raisebox{-0.13ex}{\ding{182}}--\raisebox{-0.13ex}{\ding{186}}, then read upward for definitions, examples, and functions. Continue with \raisebox{-0.13ex}{\ding{187}}--\raisebox{-0.13ex}{\ding{188}}, and finally {\protect\infoicon}. EOI~=~everyday object-based interaction; MAI~=~mid-air gesture interaction. Details in \Cref{sec:objestures}.} 
\Description{A dense, multi-part infographic diagram divided into three main sections, guiding the reader through numbered steps 1 to 7. The top section is a large table structured into five columns: Binary (B), Linear (L), Rotational (R), Nonlinear (N), and Free (F), with rows for definition, MAI and EOI examples, and function. Below the table, the area shows a sequence of physical object manipulations (e.g., tapping a bell, squeezing a sponge) paired directly with their equivalent mid-air hand gestures identified throughout the process. The bottom-left section features a 3D coordinate system representing the design space, with all three axes labeled B, L, R, N, and F. Semi-transparent planar surfaces span these axes, populated with small thumbnail images of seven applications we designed. The bottom-right section presents captures from prior work that were not considered. Together, the diagram shows how isolated examples from prior literature were synthesized into a structured framework of interchangeable and combinable object-based and mid-air interactions.}
    \label{fig:designSpace}
\end{figure*}

\Cref{fig:designSpace} (\raisebox{-0.13ex}{\ding{182}}--\raisebox{-0.13ex}{\ding{188}}) details the formation of our interaction types and design space. Grounded in Research through Design (RtD)~\cite{zimmerman_research_2007} and theory as annotation~\cite{gaver_what_2012}, our framework emerges from a reflective engagement with representative exemplars rather than an exhaustive literature survey.

In prior work, we observed that many EOIs rely on sliding~\cite{joshi_evaluation_2019, he_ubi_2023} or rotating~\cite{hettiarachchi_annexing_2016,lopes_adding_2018} \raisebox{-0.13ex}{\ding{182}}. These patterns became salient when viewed through the lens of affordance theory~\cite{gibson_ecological_2014, norman_design_2013}: a book or table edge affords guiding the hand along a linear path, while a rounded plate or cup affords rotation around a central axis. 

\citet{gong_affordance-based_2023} elicited a set of affordance-based EOIs for spatial tangible interactions, and suggested examining their usability and trade-offs relative to MAIs \raisebox{-0.13ex}{\ding{183}}, noting that one modality may be more convenient than the other (e.g., pinching may be quicker than tapping an object). This motivated us to identify corresponding MAI equivalents~\cite{hosseini_towards_2023} \raisebox{-0.13ex}{\ding{184}}. These EOI--MAI mappings can be viewed as pantomimic gestures~\cite{kendon_gesture_2004, mcneill_hand_1992}, where the same motion can be performed with or without an object.

As we continued working with examples, we observed that although squeezing~\cite{sheng_interface_2006} can sometimes trigger the same binary outcome as tapping (e.g., squeezing the object held in hand while approaching the car to open the trunk~\cite{saponas_enabling_2009}, which could also be done by pressing a button), it affords distinct experiential qualities by providing immediate feedback through changes in shape or resistance, allowing the hand to feel the action as it unfolds \raisebox{-0.13ex}{\ding{185}}. 

We then observed that the first four types entail specific kinematic constraints, whereas some EOIs involve relatively unconstrained spatial movements~\cite{hettiarachchi_annexing_2016,jain_ubi-touch_2023}. Phenomenologically~\cite{merleau-ponty_phenomenology_1945}, these are not merely unrestricted variants but distinct actions arising from motor intentionality and situated bodily knowledge \raisebox{-0.13ex}{\ding{186}}.

At the top of \Cref{fig:designSpace}, we provide a detailed characterization of the five interaction types, which have so far been identified through unimanual observations. However, many EOIs and MAIs appear in bimanual form~\cite{matulic_pensight_2020,nith_splitbody_2024,surale_experimental_2019} \raisebox{-0.13ex}{\ding{187}}. This prompted us to consider how the five types extend to bimanual interaction---each hand can independently perform any of the five types, either as an EOI or an MAI. Combinatorially, this yields a design space \raisebox{-0.13ex}{\ding{188}}: unimanual interactions form two 1D regions---EOIs \raisebox{-0.13ex}{\ding{172}} and MAIs \raisebox{-0.13ex}{\ding{173}}---each comprising five types, while bimanual interactions contribute three additional 2D regions: EOI~$\times$~MAI \raisebox{-0.13ex}{\ding{174}}, EOI~$\times$~EOI \raisebox{-0.13ex}{\ding{175}}, and MAI~$\times$~MAI \raisebox{-0.13ex}{\ding{176}}. Each bimanual region contains 25 possibilities ($5 \times 5$). Together, these five regions yield 85 possible interactions. Evaluating all combinations is beyond the scope of this paper; therefore, we selected seven instances spanning all five types for an exploratory user study and case studies, with their positions indicated in the design space. Additional potential applications are illustrated in \Cref{fig:designSpace-potential}.

We focus on independent whole-hand movements, which enable us to identify recurring patterns across EOIs and MAIs and the correspondences between them. In contrast, interactions involving one hand acting on the other hand~\cite{pei_hand_2022}, on an object held in that hand~\cite{zhou_gripmarks_2020}, or fine-grained finger movements~\cite{joshi_transferable_2023,gil_proptype_2025} tend to be highly specialized, with limited shared affordances across objects and mid-air gestures, thus falling outside our scope {\infoicon}. 

\Cref{tab:prior-work} situates representative prior work through the lens of Objestures. The design space remains open~\cite{lupetti_making_2024,lyu_unbounded_2026}, and additional types may be incorporated as designers encounter further compelling examples. 

\begin{table*}
\footnotesize
\caption{Interaction Types and Modalities Observed in Representative Prior Work}
\Description{A comprehensive comparison matrix structured as a data table. The left-most column contains a vertical list of prior research works, culminating in a bolded final row for the Objestures. The main column headers categorize interactions into five types (Binary, Linear, Rotational, Nonlinear, and Free) and modalities (Unimanual, Bimanual, and multiple types). Beneath these headers, sub-columns are visually distinguished by small geometric icons. The grid's cells are populated with checkmarks indicating full presence, short dashes indicating partial presence, or left blank. The far-right column, ``Realization,'' provides brief text descriptions of the hardware or study methodologies used for each row. Visually, the table demonstrates the capabilities of prior work through scattered checkmarks, contrasted by the bottom ``Objestures'' row, which features a continuous, unbroken line of checkmarks across the columns. A legend at the bottom defines the geometric symbols and technical acronyms.}
\vspace{-1em}
\label{tab:prior-work}
\rowcolors{10}{gray!3}{white} 

\begin{tabular*}{\textwidth}{@{\extracolsep{\fill}}
l
*{10}{c}
c      
cc
ccc
c      
cc
l
@{}}

\toprule

& \multicolumn{2}{c}{Binary}
& \multicolumn{2}{c}{Linear}
& \multicolumn{2}{c}{Rotational}
& \multicolumn{2}{c}{Nonlinear}
& \multicolumn{2}{c}{Free}
&
& \multicolumn{2}{c}{Unimanual}
& \multicolumn{3}{@{}c@{}}{Bimanual}
&
& \multirow{2}{*}[-0.28em]{{n}~\textcolor{MyBlue}{$\blacksquare$}}
& \multirow{2}{*}[-0.28em]{{n}~\textcolor{MyGreen}{$\blacktriangle$}}
& \multirow{2}{*}[-0.28em]{Realization} \\

\cmidrule(lr){2-3}
\cmidrule(lr){4-5}
\cmidrule(lr){6-7}
\cmidrule(lr){8-9}
\cmidrule(lr){10-11}
\cmidrule(lr){13-14}
\cmidrule(lr){15-17}

& \textcolor{MyBlue}{$\blacksquare$} & \textcolor{MyGreen}{$\blacktriangle$}
& \textcolor{MyBlue}{$\blacksquare$} & \textcolor{MyGreen}{$\blacktriangle$}
& \textcolor{MyBlue}{$\blacksquare$} & \textcolor{MyGreen}{$\blacktriangle$}
& \textcolor{MyBlue}{$\blacksquare$} & \textcolor{MyGreen}{$\blacktriangle$}
& \textcolor{MyBlue}{$\blacksquare$} & \textcolor{MyGreen}{$\blacktriangle$}
&
& \textcolor{MyBlue}{$\blacksquare$} & \textcolor{MyGreen}{$\blacktriangle$}
& \textcolor{MyBlue}{$\blacksquare$}+\textcolor{MyBlue}{$\blacksquare$}
& \textcolor{MyGreen}{$\blacktriangle$}+\textcolor{MyGreen}{$\blacktriangle$}
& \textcolor{MyBlue}{$\blacksquare$}+\textcolor{MyGreen}{$\blacktriangle$}
&
& & \\

\midrule

\mbox{Ad Hoc UI~\cite{du_opportunistic_2022}} 
& $\checkmark$ & & $\checkmark$ & & $\checkmark$ & & & & $\checkmark$ & 
& & $\checkmark$ & & & & & 
& $\checkmark$ & 
& \mbox{Phone camera} \\ 

\mbox{Affordance-Based~\cite{gong_affordance-based_2023}} 
& $\checkmark$ & & $\checkmark$ & & $\checkmark$ & & $\checkmark$ & & $\checkmark$ & 
& & $\checkmark$ & & $\checkmark$ & & & 
 & $\checkmark$ & 
& \mbox{Elicitation study~+~WoZ} \\ 

\mbox{Always-Available~\cite{saponas_enabling_2009}} 
& $\checkmark$ & $\checkmark$ & & & & & --- & --- & & 
& & $\checkmark$ & $\checkmark$ & $\checkmark$ & $\checkmark$ & --- & 
 & $\checkmark$ & 
& \mbox{EMG} \\ 

\mbox{Annexing Reality~\cite{hettiarachchi_annexing_2016}} 
& & & & & $\checkmark$ & & & & $\checkmark$ & 
& & $\checkmark$ & & $\checkmark$ & & & 
 & $\checkmark$ & 
& \mbox{Kinect} \\ 

\mbox{Consensus Set~\cite{hosseini_towards_2023}} 
& & $\checkmark$ & & $\checkmark$ & & $\checkmark$ & & $\checkmark$ & & $\checkmark$ 
& & & $\checkmark$ & & $\checkmark$ & 
& & & $\checkmark$ 
& \mbox{Literature review} \\ 

\mbox{EMS Haptics~\cite{lopes_adding_2018}}
& ---&$\checkmark$ 
& ---& $\checkmark$
&$\checkmark$ &--- 
&$\checkmark$ &$\checkmark$ 
&--- &--- 
& 
&$\checkmark$ &$\checkmark$ 
&$\checkmark$ &--- &--- 
& 
&$\checkmark$ 
&$\checkmark$ 
& \mbox{HMD + EMS}\\ 

\mbox{Haptic AR~\cite{bhatia_augmenting_2024}} 
& $\checkmark$ & & $\checkmark$ & & & & $\checkmark$ & & & 
& & $\checkmark$ & & & & & 
& $\checkmark$ & 
& \mbox{Literature review~+~WoZ} \\ 

\mbox{Haptics at Home~\cite{fang_vr_2023}} 
& $\checkmark$ & & $\checkmark$ & & $\checkmark$ & & $\checkmark$ & & $\checkmark$ & 
& & $\checkmark$ & & $\checkmark$ & & & 
& $\checkmark$ & 
& \mbox{HMD} \\ 

\mbox{iCon~\cite{cheng_icon_2010}} 
& $\checkmark$ & & & & $\checkmark$ & & & & $\checkmark$ & 
& & $\checkmark$ & & & & & 
 & $\checkmark$ & 
& \mbox{Webcam~+~pattern stickers} \\ 

\mbox{Input at the Edge~\cite{joshi_evaluation_2019}}
&$\checkmark$ & 
&$\checkmark$ & 
& & 
& & 
& & 
& 
&$\checkmark$ & 
& ---& &--- 
& 
&--- 
& 
& \mbox{Projector~+~Vicon}\\ 

\mbox{Mobile Haptics~\cite{stellmacher_exploring_2024}} 
& $\checkmark$ & & $\checkmark$ & & & & $\checkmark$ & & $\checkmark$ & 
& & $\checkmark$ & & $\checkmark$ & & $\checkmark$ 
& & & 
& \mbox{Elicitation study} \\ 

\mbox{Mode-Switching~\cite{surale_experimental_2019}} 
& --- & $\checkmark$
& & $\checkmark$
& & ---
& & 
& & ---
& 
& &$\checkmark$ 
& &$\checkmark$ &--- 
& 
& 
& $\checkmark$
& \mbox{HMD~+~Leap Motion~+~FSR}\\ 

\mbox{Pan-and-Zoom~\cite{nancel_mid-air_2011}} 
& $\checkmark$ & $\checkmark$ & $\checkmark$ & $\checkmark$ & $\checkmark$ & $\checkmark$ & & & $\checkmark$ & $\checkmark$ 
& & $\checkmark$ & $\checkmark$ & $\checkmark$ & $\checkmark$ & $\checkmark$ 
& & $\checkmark$ & $\checkmark$ 
& \mbox{Vicon~+~phone~+~mouse} \\ 

\mbox{Passive Tangibles~\cite{drogemuller_turning_2021}} 
& & & $\checkmark$ & & $\checkmark$ & & & & & 
& & $\checkmark$ & & & & & 
& $\checkmark$ & 
& \mbox{HMD~+~markers band} \\ 

\mbox{PenSight~\cite{matulic_pensight_2020}} 
& $\checkmark$ & $\checkmark$ & $\checkmark$ & $\checkmark$ & --- & & & & $\checkmark$ & --- 
& & $\checkmark$ & $\checkmark$ & $\checkmark$ & $\checkmark$ & $\checkmark$ 
& & & $\checkmark$ 
& \mbox{Pen-top camera} \\ 

\mbox{Pen~+~Mid-Air~\cite{aslan_pen_2018}}
& $\checkmark$ & $\checkmark$ & $\checkmark$ & $\checkmark$ & & $\checkmark$ & & --- & $\checkmark$ & $\checkmark$ 
& & & & & & $\checkmark$ 
& & & $\checkmark$ 
& \mbox{Elicitation study} \\ 

\mbox{Pen~+~Touch~\cite{hinckley_pen_2010}} 
& $\checkmark$ & --- & $\checkmark$ & --- & --- & --- & & & $\checkmark$ & --- 
& & $\checkmark$ & --- & & & --- 
& & & $\checkmark$ 
& \mbox{Tabletop computer~+~pen} \\ 

\mbox{Proxy Sculpting~\cite{sheng_interface_2006}}
&$\checkmark$ &$\checkmark$ 
&$\checkmark$ &$\checkmark$ 
&$\checkmark$ &$\checkmark$ 
&$\checkmark$ & 
&$\checkmark$ &$\checkmark$ 
& 
&$\checkmark$ &--- 
&$\checkmark$ &$\checkmark$ &$\checkmark$ 
& 
&$\checkmark$ 
&$\checkmark$ 
& \mbox{Vicon}\\ 

\mbox{SplitBody~\cite{nith_splitbody_2024}} 
& $\checkmark$ & $\checkmark$ & --- & --- & $\checkmark$ & --- & & & $\checkmark$ & $\checkmark$ 
& & $\checkmark$ & $\checkmark$ & $\checkmark$ & & $\checkmark$ 
& & $\checkmark$ & --- 
& \mbox{EMS}\\

\mbox{TabletInVR~\cite{surale_tabletinvr_2019}} 
& $\checkmark$ & $\checkmark$ & & --- & & --- & & & $\checkmark$ & --- 
& & $\checkmark$ & $\checkmark$ & $\checkmark$ & & $\checkmark$ 
& & & $\checkmark$ 
& \mbox{HMD~+~tracker} \\ 

\mbox{Teachable Reality~\cite{monteiro_teachable_2023}} 
& $\checkmark$ & & $\checkmark$ & & $\checkmark$ & & $\checkmark$ & & $\checkmark$ & 
& & $\checkmark$ & & $\checkmark$ & & & 
& $\checkmark$ & 
& \mbox{Phone camera} \\ 

\mbox{Ubi Edge~\cite{he_ubi_2023}}
&$\checkmark$ & 
&$\checkmark$ & 
&--- & 
& & 
& & 
& 
&$\checkmark$ & 
&--- & &--- 
& 
&$\checkmark$ 
& 
& \mbox{HMD~+~LiDAR}\\ 

\mbox{Ubi-TOUCH~\cite{jain_ubi-touch_2023}}
&$\checkmark$ &$\checkmark$ 
&$\checkmark$ & 
&$\checkmark$ & 
&$\checkmark$ & 
&$\checkmark$ & 
& 
&$\checkmark$ & 
& & & 
& 
&$\checkmark$ 
& 
& \mbox{HMD + depth camera}\\ 
\midrule
\addlinespace[0.3em]
\hiderowcolors
\mbox{\textbf{Objestures}} 
& $\pmb{\checkmark}$ & $\pmb{\checkmark}$ & $\pmb{\checkmark}$ & $\pmb{\checkmark}$ & $\pmb{\checkmark}$ & $\pmb{\checkmark}$ & $\pmb{\checkmark}$ & $\pmb{\checkmark}$ & $\pmb{\checkmark}$ & $\pmb{\checkmark}$ 
& & $\pmb{\checkmark}$ & $\pmb{\checkmark}$ & $\pmb{\checkmark}$ & $\pmb{\checkmark}$ & $\pmb{\checkmark}$ 
& & $\pmb{\checkmark}$ & $\pmb{\checkmark}$ 
& \mbox{HMD} \\

\bottomrule
\end{tabular*}

{{\vspace{0.2em}
\footnotesize
\mbox{
\textcolor{MyBlue}{$\blacksquare$} = EOIs, 
\textcolor{MyGreen}{$\blacktriangle$} = MAIs;\ \ \ 
{n}~X = multiple types of X interactions;\ \ \  
$\checkmark$ = present,  
\textbf{--} = partially present or implied};\ \ \ 
WoZ = Wizard of Oz experiment, 
EMG = electromyography, 
EMS = electrical muscle stimulation, 
HMD = head-mounted display, 
FSR = force-sensitive resistor, 
LiDAR = light detection and ranging 
}}
\end{table*}

\paragraph{Abstraction and Productive Ambiguity}
Bimanual theory~\cite{guiard_asymmetric_1987,buxton_study_1986} emphasizes asymmetric roles (dominant vs.~non-dominant hands), but our design space abstracts away from rigid hand assignment toward shared motion patterns. This reflects an RtD stance: the framework specifies combinable possibilities, while role assignment remains an interpretive layer for practitioners (e.g., Rotation can be enacted coarsely or finely, and designers may map these nuances to hand roles depending on context). Since our types are defined by high-level patterns rather than strict typologies, some interactions may be ambiguous. For example, interacting with a spray-can trigger can be interpreted as Rotational (hinged pivoting) or Nonlinear (spring-loaded compression). Such ambiguity is productive rather than limiting~\cite{gaver_ambiguity_2003}, allowing practitioners to foreground the qualities salient to their intended experience.

\paragraph{Implementation}
Details are provided in \Cref{appen:implementation-details}, as our goal is to prototype design ideas rather than build a complete system. Prior EOI work has largely relied on object tracking~\cite{monteiro_teachable_2023, sheng_interface_2006, furumoto_midair_2021, hettiarachchi_annexing_2016}, potentially reinforcing a divide from MAIs that use hand tracking. Taking a holistic view, we note that hands are central to both modalities and that object movement can often be inferred from hand motion (e.g., when turning a coffee cup, the hand and object move together). Hence, we rely solely on the headset's built-in hand tracking for both EOIs and MAIs. Our approach has limitations in handling occlusion and inertia: tracking may degrade when fingers are partially occluded, and an object may continue moving after release. We thus assume hands remain visible and in contact with objects during interaction. As in prior work~\cite{hinckley_pen_2010}, this suffices for prototyping, while robust handling remains future work.

\section{User Study}\label{sec:study1}
This exploratory study evaluates the usefulness of \oj{} in supporting basic spatial manipulations. Pilot studies were conducted with two lab members to refine the procedure and details (e.g., the distance and size of the target cubes).

\subsection{Participants and Apparatus}\label{sec:participants_apparatus}
We recruited 12 participants (6 females, 6 males) aged 22--32 (M = 25.3, SD = 3.0). Seven had prior XR experience: six used XR equipment a few times a year, and one used it weekly. Among these seven, the median familiarity rating with XR functionality and applications was 5 (IQR = 4--6) on a 7-point scale, with all ratings 4 and above. Five of the seven had used MR features (Passthrough/Seethrough). Participants reported using XR for gaming, virtual classrooms/labs, artistic experiences, and research. All but one were right-handed, and all had normal or corrected-to-normal vision. Participants received compensation for their involvement.

The study utilized the Passthrough feature of the Meta Quest~3 (build~63.0) via Quest Link on a Windows~10 PC equipped with a GeForce GTX 1070 GPU, using Unity (2022.3.14f1) with Oculus Integration (v57). The system supported both right- and left-handed users. The setup included a studio light for enhanced illumination, a desk (65~cm tabletop height) with a chair, a hardcover book (26~cm edge length), and a disposable coffee cup with a lid (9~cm top diameter).

\subsection{Design}\label{sec:des} 
\begin{figure*}
    \centering
    \includegraphics[width=\textwidth]{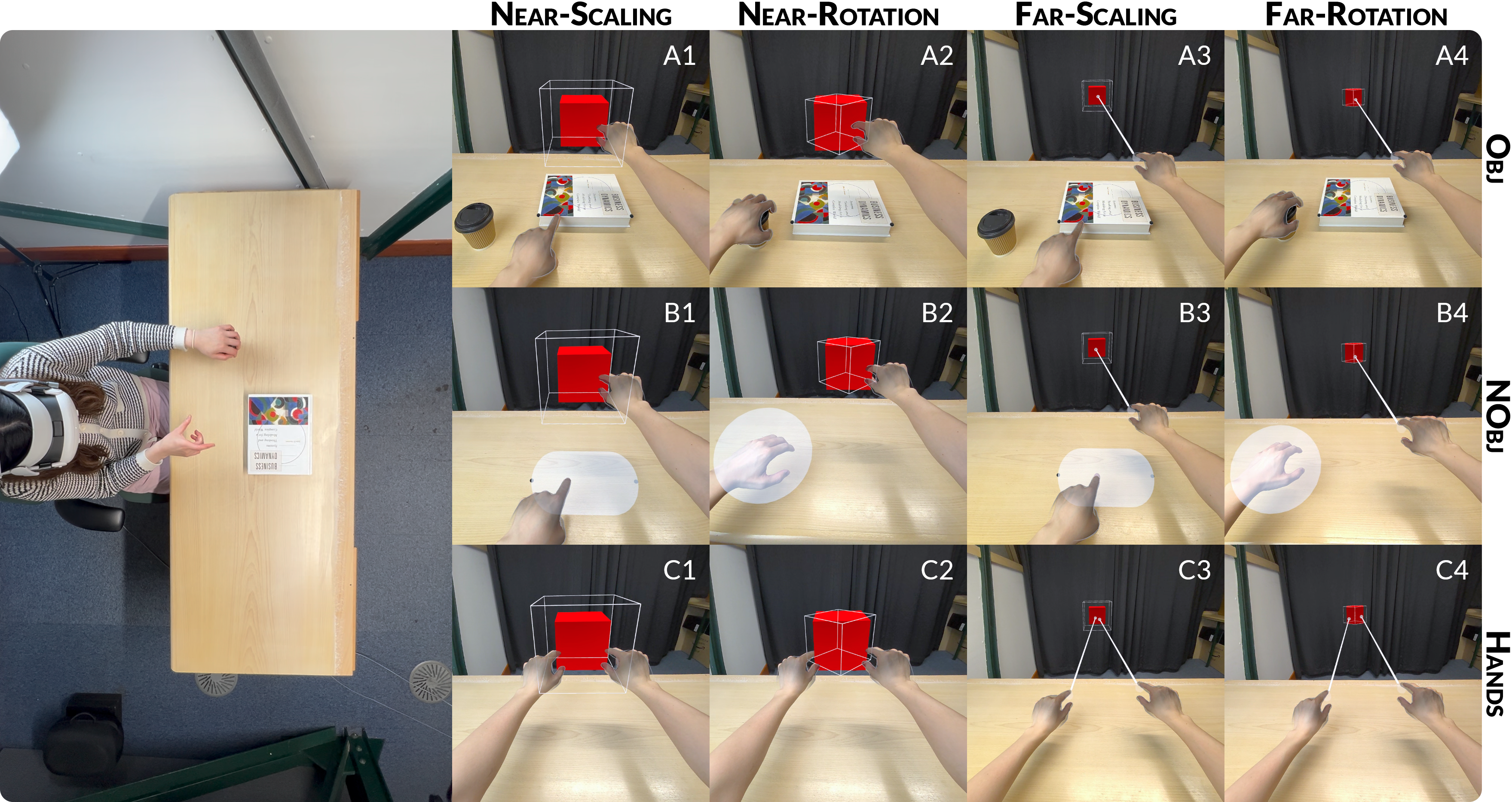}
    \caption{Overview of study setup and conditions. The image on the left shows the study environment. The sub-figures (A1--C4) on the right illustrate the three Approaches (rows: \textsc{Obj}, \textsc{NObj}, \textsc{Hands}) across four Task~$\times$~Distance conditions (columns: \textsc{Near-Scaling}, \textsc{Near-Rotation}, \textsc{Far-Scaling}, \textsc{Far-Rotation}). Each sub-figure shows the moment immediately before the cube was pinch-selected to initiate manipulation. Users manipulate the cubes to match the target angle or scale shown in the white wireframe. Refer to \Cref{sec:des} and the accompanying video for details.}
\Description{The figure is composed of two parts: on the left, a large photo shows a top-down view of a study participant sitting at a table, wearing a head-mounted display, with one hand placed on a book and the other hand performing a pinching gesture in mid-air. A disposable coffee cup is placed next to the book. The surrounding environment is a small, enclosed lab space with a wooden table in front of the participant. On the right, there are 12 subfigures arranged in a 3×4 grid. Each subfigure is labeled (A1--C4) and shows a different interaction. The interactions involve manipulating a red 3D cube across different Distances and Approaches. From top to bottom: the first row (A1--A4) represents Obj, where one hand interacts with the cube and the other hand interacts with a physical object (cup or book). The second row (B1--B4) represents NObj, where one hand interacts with the cube and the other hand interacts in translucent interaction zones (shown as spheres or capsules). The third row (C1--C4) represents Hands, where both hands directly manipulate the cube. From left to right: the first column (A1, B1, C1) involves Near-Scaling. The second column (A2, B2, C2) focuses on Near-Rotation. The third column (A3, B3, C3) involves Far-Scaling. The fourth column (A4, B4, C4) demonstrates Far-Rotation.}
    \label{fig:study1}
\end{figure*}

As shown in \Cref{fig:study1}, the experiment uses a 3~$\times$~2~$\times$~2 within-subjects design with the following factors and levels: \textit{Approach} (\textsc{Hands}, \textsc{Obj}, \textsc{NObj}), \textit{Task} (\textsc{Scaling}, \textsc{Rotation}), and \textit{Distance} (\textsc{Near}, \textsc{Far}). The performance measures are \textit{Error}, \textit{Movement}, and \textit{Time}.

\mySubsub{Independent Variables}
\mylabel{Approach} In line with prior work~\cite{song_handle_2012, pei_hand_2022, song_hotgestures_2023}, the headset's native freehand interaction serves as the baseline (\textsc{Hands}), involving direct manipulation (for \textsc{Near}) and ray interaction (for \textsc{Far}). Controllers are excluded---despite their prevalence in VR---because they can be inconvenient~\cite{stellmacher_exploring_2024} and cumbersome~\cite{zhou_gripmarks_2020} in MR, as holding them hinders direct interaction with the physical environment.

Since \oj{} can be performed either with or without physical objects (\Cref{fig:designSpace}), our comparison includes both. For \textsc{Obj}, the dominant hand performs a Binary MAI through pinching to select the cube either directly (\textsc{Near}) or via a ray (\textsc{Far}), while the non-dominant hand performs either a Linear EOI (\textsc{Scaling}) by sliding along the book edge or a Rotational EOI (\textsc{Rotation}) by rotating the cup, which participants grip from the top for more accurate hand tracking~\cite{schlesinger_mechanische_1919, mackenzie_grasping_1994, aslan_pen_2018, dim_designing_2016}. For \textsc{NObj}, the dominant hand's MAI (cube selection) remains the same as in \textsc{Obj}, but the non-dominant hand's EOIs are replaced by MAIs within translucent virtual zones: a capsule 26~cm long (matching the book edge) and 15~cm in diameter for \textsc{Scaling}, and a sphere 20~cm in diameter for \textsc{Rotation}. 

\medskip
\mylabel{Task} We select two common 3D manipulations: \textsc{Scaling} and \textsc{Rotation}~\cite{bowman_3d_2004}, which involve Linear, Rotational, and Binary interactions. Free and Nonlinear interactions are hence examined in the case studies (\Cref{sec:study2}) for their experiential quality, although we acknowledge the lack of their performance evaluation as a limitation.

We use white wireframes~\cite{song_handle_2012} to represent target scales or angles. Each Task starts with a red virtual cube (center-fixed; 15~cm side length, 0° rotation) that turns green upon selection. \textsc{Scaling} requires resizing the cube to match a target side length ranging randomly from 5--12~cm (scale down) or 18--25~cm (scale up). \textsc{Rotation} involves horizontally rotating the cube to match a target angle ranging randomly from 7.5° to 45° (rotate left) or $-$7.5° to $-$45° (rotate right). Our testing and pilot studies informed these ranges: \textsc{Scaling} bounds ensure the target is neither too small nor too large to be comfortably matched by the cube, while guaranteeing a $\ge$~3~cm noticeable difference from the 15~cm default; for \textsc{Rotation}, an absolute 7.5° minimum ensures discernibility at \textsc{Far}, while the absolute 45° maximum accounts for the cube's symmetry (e.g., a 60° target is visually identical to a $-$30° target).

\medskip
\mylabel{Distance} A \textsc{Near} cube is located within a reachable area above the desk, with its center positioned randomly inside an invisible 20~cm cubic region centered 40~cm in front of the headset's line of sight and 30~cm above the desk. A \textsc{Far} cube is placed in an out-of-reach area, with its center positioned randomly inside an invisible 70~cm (3.5~$\times$~\textsc{Near}) cubic region, centered 140~cm (3.5~$\times$~\textsc{Near}) from the headset at the same height as the \textsc{Near} region.

\mySubsub{Dependent Variables}
For \textsc{Scaling}, \textit{Error} is defined as $|1 - \frac{\text{manipulated scale}}{\text{target scale}}|$ in percentage (\%), where scale is measured by the cube's side length. For \textsc{Rotation}, Error is defined as $|\text{manipulated angle} - \text{target angle}|$, measured in degrees (°). For both Tasks, \textit{Movement} is the total distance moved by both arms during a trial, calculated as the cumulative translational distance of the wrists in meters (m); \textit{Time} is the duration in seconds (s) from when the cube appears to when it is released by the user after manipulation. We formulated the following hypotheses regarding the main effects of {Approach}:
\begin{itemize}[label=\textbullet,leftmargin=*]
    \setlength\itemsep{0em}
    \item {H1:} We expect \textsc{Obj} to yield lower Error than \textsc{NObj} due to physical guidance. Neither \textsc{Obj} nor \textsc{NObj} is expected to perform significantly worse than \textsc{Hands}.
    \item {H2:} We expect both \textsc{Obj} and \textsc{NObj} to reduce Movement relative to \textsc{Hands} as they separate the roles of targeting (dominant hand) and manipulation (non-dominant hand). We expect no significant difference between \textsc{Obj} and \textsc{NObj} due to similar interaction logic.
    \item {H3:} We expect \textsc{Obj} and \textsc{NObj} to show Time performance comparable to, or faster than, \textsc{Hands}, as separating hand roles may reduce coordination demands.
\end{itemize}

\mySubsub{Trials}
Each trial is one of four types: \textsc{Near}-\textsc{Scaling} (NS), \textsc{Near}-\textsc{Rotation} (NR), \textsc{Far}-\textsc{Scaling} (FS), or \textsc{Far}-\textsc{Rotation} (FR). Two consecutive trials of the same type (e.g., 2~NS) form a pair. To minimize clutching (i.e., the need to reset hand positions), each pair alternates between opposite manipulation directions (e.g., scale-up then scale-down). Four pairs (one per type) in random order form a block (e.g., 2~NS $\rightarrow$ 2~FR $\rightarrow$ 2~FS $\rightarrow$ 2~NR). This design ensures equal exposure to each type, preventing the overrepresentation of any single type (e.g., 4~NS), while introducing randomness to counteract within-block order effects. 

For each Approach, participants complete two training blocks followed by three test blocks from which data are included in the statistical analysis. To mitigate order effects, the three Approaches are fully counterbalanced, with two participants assigned to each of the six permutations. Aside from training, the design yields 12 participants $\times$ 3 Approaches $\times$ 3 blocks $\times$ 8 trials/block = 864 trials.

\subsection{Procedure}\label{sec:pro}
Participants reviewed the information sheet, signed a consent form, and completed a demographic questionnaire. They then proceeded through the three Approaches in their assigned order.

For each Approach, participants began with training blocks where they were instructed on its use and practiced with it. During the subsequent test blocks, they were asked to perform as quickly and accurately as possible without undue pressure. Each trial in these blocks required participants to pinch to select the cube and maintain this selection while manipulating it. Once satisfied, they released the pinch, upon which Error, Movement, and Time were recorded; the cube disappeared, and the next one appeared. Between phases and Approaches, participants could take a short break and adjust or remove the headset if needed.

After completing an Approach, participants filled out the NASA-TLX~\cite{hart_nasa-task_2006} and UEQ-S~\cite{laugwitz_construction_2008} questionnaires, provided feedback on the Approach's positive and negative qualities, and shared additional verbal or written comments. Based on this feedback, the experimenter followed up iteratively to clarify (e.g., ``Was it challenging to rotate your hand to the left?'').

After completing all Approaches, participants ranked them by preference and explained their reasoning. The experimenter conducted a final brief interview to explore these preferences. The mean duration of the study was 51~min 58~s (SD = 10~min 51~s).

\subsection{Results: Performance}\label{sec:performance}
We observed that Oculus Integration sometimes resulted in false triggers for users with small hands, particularly when a slight pinch was interpreted as an intentional input, causing the cube to be released shortly after the hand touched it without significant manipulation. Therefore, we excluded trials with interaction time~<~0.5~s, scale change~<~0.001~m, or rotation angle~<~1°, totaling 4.51\% of the data. Additionally, extreme outliers were excluded: \textsc{Scaling} Error > 50\%, \textsc{Rotation} Error~>~10°, Movement~>~5~m, or Time~>~15~s, totaling 1.85\% of the data. This resulted in 809 valid data points out of 864 recorded, with each subject having 2 to 6 data points per condition (M = 5.62, SD = 0.79) for ANOVA. 

\begin{figure*}
\centering
\includegraphics[width=\textwidth]{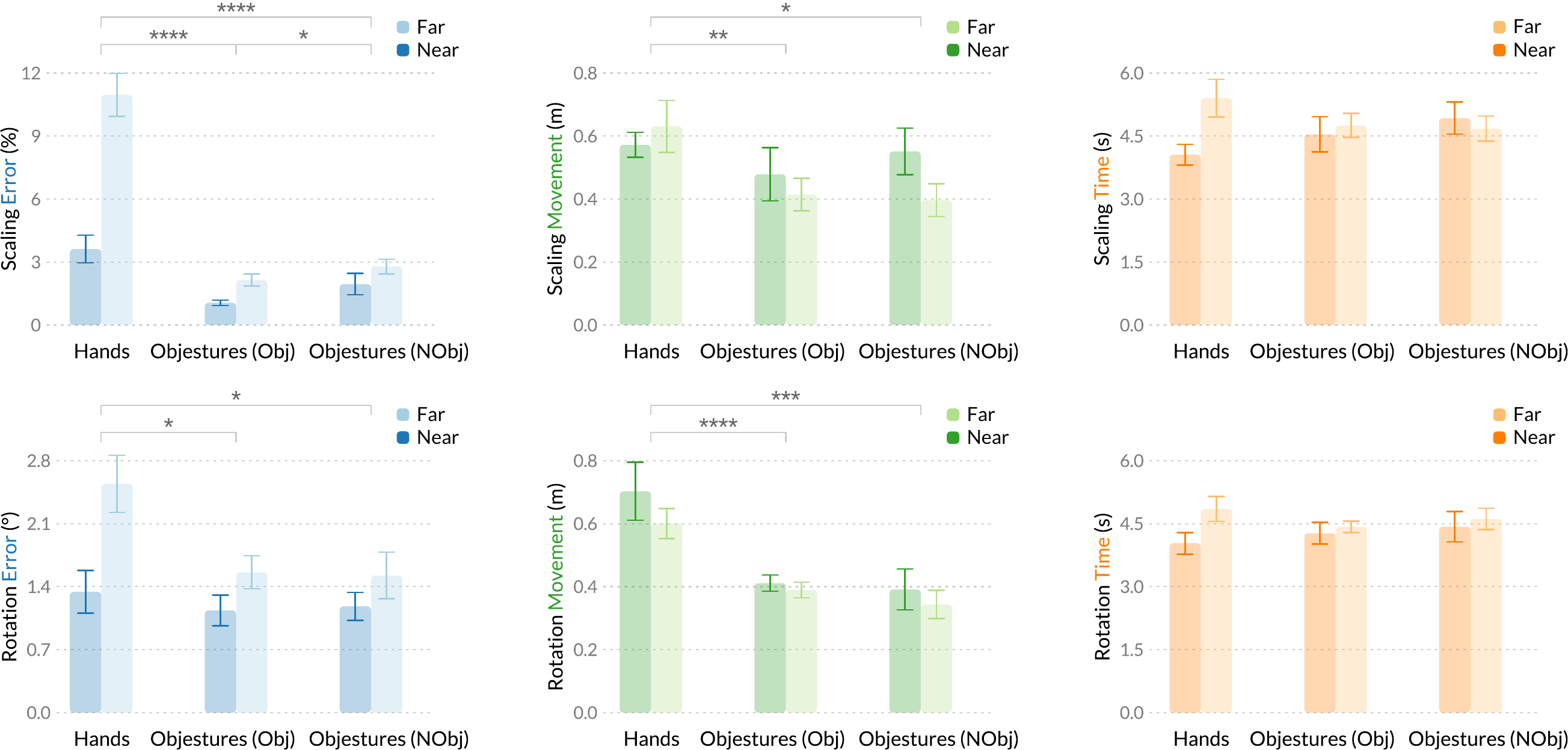}
\caption{Plots of performance metrics (Error, Movement, Time) for \textsc{Scaling} and \textsc{Rotation} Tasks across different Approaches (\textsc{Hands}, \textsc{Obj}, \textsc{NObj}) and Distances (\textsc{Near}, \textsc{Far}). Error bars represent standard errors. Significant differences (main effects of Approach only) are indicated by * (p < 0.05), ** (p < 0.01), *** (p < 0.001), and **** (p < 0.0001); pairwise differences at individual Distance levels are not implied. See \Cref{sec:performance} for details.}
\Description{A two-by-three grid of bar charts presenting data across two rows and three columns. The top row displays charts for ``Scaling'' metrics, while the bottom row displays charts for ``Rotation'' metrics. From left to right, the columns are color-coded by metric: ``Error'' (blue), ``Movement'' (green), and ``Time'' (orange). Every individual chart shares an identical x-axis structure featuring three main categories: ``Hands'', ``Objestures (Obj)'', and ``Objestures (NObj)''. Within each of these categories, two adjacent vertical bars are plotted side-by-side to represent Distance, distinguished by color intensity: a lighter shade for ``Far'' and a darker shade for ``Near.'' The charts in the ``Error'' (blue) and ``Movement'' (green) columns feature horizontal brackets with varying numbers of asterisks indicating statistical significance. Specifically, for ``Scaling Error'', both Obj and NObj are significantly lower than Hands (four asterisks each), with Obj also being significantly lower than NObj (one asterisk). For ``Scaling Movement'', both Obj (two asterisks) and NObj (one asterisk) are significantly lower than Hands. For ``Rotation Error'', both Obj and NObj are significantly lower than Hands (one asterisk each). For ``Rotation Movement'', both Obj (four asterisks) and NObj (three asterisks) are significantly lower than Hands. The ``Time'' (orange) charts do not indicate any statistical significance.}
\label{fig:performance}
\end{figure*}

We analyzed the \textsc{Scaling} and \textsc{Rotation} Tasks separately to examine the effects of {Approach} and {Distance} on {Error}, {Movement}, and {Time} (\Cref{fig:performance}). We report generalized eta-squared (\(\eta^2_G\)) effect sizes for ANOVA to allow comparisons across designs \mbox{\cite{olejnik_generalized_2003, bakeman_recommended_2005}}. Post-hoc power analyses were conducted for the main effects of Approach using G*Power~\cite{faul_gpower_2007}, with $\alpha = 0.05$, a sample size of 12, and effect sizes calculated by partial eta squared. The achieved power ($1 - \beta$) is reported alongside the F-statistics. Post hoc pairwise comparisons were performed using t-tests on estimated marginal means (EMMs) with Bonferroni adjustment. Means are tabulated for quick reference (see \Cref{tab:means-table-all}). Due to the balanced design, EMMs are mathematically identical to the specific means used in each analysis (i.e., log-transformed or raw) and are thus omitted. For readability, long statistics are placed in the footnote. 

\mySubsub{Error}
For both \textsc{Scaling} and \textsc{Rotation}, after fitting two-way repeated measures ANOVAs, the Shapiro-Wilk tests on the residuals indicated violations of the normality assumption ($W_{\text{\textsc{Scaling}}} = 0.934$, $p < 0.001$; $W_{\text{\textsc{Rotation}}} = 0.931$, $p < 0.001$). The Kolmogorov-Smirnov tests showed that the data for both Tasks followed log-normal distributions ($D_{\text{\textsc{Scaling}}} = 0.081$, $p = 0.695$; $D_{\text{\textsc{Rotation}}} = 0.063$, $p = 0.916$). Log-transforming the data resolved normality issues ($W_{\text{\textsc{Scaling}}} = 0.988$, $p = 0.715$; $W_{\text{\textsc{Rotation}}} = 0.991$, $p = 0.905$). Mauchly's tests confirmed sphericity for Approach ($W_{\text{\textsc{Scaling}}} = 0.864$, $p = 0.481$; $W_{\text{\textsc{Rotation}}} = 0.956$, $p = 0.800$) and Approach \(\times\) Distance ($W_{\text{\textsc{Scaling}}} = 0.867$, $p = 0.490$; $W_{\text{\textsc{Rotation}}} = 0.737$, $p = 0.218$). All results presented below are based on log-transformed data.

The analysis revealed significant main effects for Distance, with \textsc{Near} showing significantly lower Error than \textsc{Far} for both \textsc{Scaling} ($F_{1, 11} = 43.77$, $p < 0.0001$, $\eta^2_G = 0.376$) and \textsc{Rotation} ($F_{1, 11} = 17.42$, $p < 0.01$, $\eta^2_G = 0.156$). 

There were significant main effects of Approach for both \textsc{Scaling} ($F_{2, 22} = 49.55$, $p < 0.0001$, $\eta^2_G = 0.562$, $1 - \beta = 1.00$) and \textsc{Rotation} ($F_{2, 22} = 6.03$, $p < 0.01$, $\eta^2_G = 0.092$, $1 - \beta = 1.00$). Post hoc analyses indicated that both \textsc{Obj} and \textsc{NObj} significantly reduced Error compared to \textsc{Hands} for \textsc{Scaling}\footnote{$\text{Est} = 1.394$, $\text{SE} = 0.168$, $t(11) = 8.28$, $p < 0.0001$;\\ $\phantom{{}^{5}}\text{Est} = 1.018$, $\text{SE} = 0.140$, $t(11) = 7.30$, $p < 0.0001$} and \textsc{Rotation}\footnote{$\text{Est} = 0.328$, $\text{SE} = 0.098$, $t(11) = 3.35$, $p < 0.05$;\\ $\phantom{{}^{5}}\text{Est} = 0.332$, $\text{SE} = 0.113$, $t(11) = 2.93$, $p < 0.05$}. Additionally, for \textsc{Scaling}, \textsc{Obj} showed significantly lower Error than \textsc{NObj} ($\text{Est} = 0.376$, $\text{SE} = 0.123$, $t(11) = 3.05$, $p < 0.05$). For \textsc{Rotation}, there was no significant difference between \textsc{Obj} and \textsc{NObj} in Error ($\text{Est} = 0.004$, $\text{SE} = 0.117$, $t(11) = 0.03$, $p = 1.000$). 

For \textsc{Scaling}, there was a significant interaction effect between Distance and Approach ($F_{2, 22} = 5.09, p < 0.05, \eta^2_G = 0.085$). Post hoc analyses revealed that for \textsc{Near}, \textsc{Obj} had significantly lower Error than \textsc{Hands} ($\text{Est} = 1.103$, $\text{SE} = 0.207$, $t(11) = 5.32$, $p < 0.001$), with no significant difference between \textsc{NObj} and \textsc{Hands} ($\text{Est} = 0.651$, $\text{SE} = 0.250$, $t(11) = 2.61$, $p = 0.073$) or between \textsc{NObj} and \textsc{Obj} ($\text{Est} = 0.452$, $\text{SE} = 0.160$, $t(11) = 2.82$, $p = 0.0502$). For \textsc{Far}, both \textsc{Obj} ($\text{Est} = 1.685$, $\text{SE} = 0.191$, $t(11) = 8.82$, $p < 0.0001$) and \textsc{NObj} ($\text{Est} = 1.385$, $\text{SE} = 0.129$, $t(11) = 10.72$, $p < 0.0001$) had significantly lower Error than \textsc{Hands}, with no significant difference between \textsc{NObj} and \textsc{Obj} ($\text{Est}  = 0.300$, $\text{SE} = 0.174$, $t(11) = 1.72$, $p = 0.338$).
For \textsc{Rotation}, there was no significant interaction effect between Distance and Approach ($F_{2, 22} = 1.97, p = 0.174, \eta^2_G = 0.036$). 

\mySubsub{Movement}
We fitted two-way repeated measures ANOVAs for \textsc{Scaling} and \textsc{Rotation}. Shapiro-Wilk tests on the residuals indicated non-normality ($W_{\text{\textsc{Scaling}}} = 0.886$, $p < 0.0001$; $W_{\text{\textsc{Rotation}}} = 0.847$, $p < 0.0001$). Kolmogorov-Smirnov tests showed that the data followed log-normal distributions ($D_{\text{\textsc{Scaling}}} = 0.080$, $p = 0.712$; $D_{\text{\textsc{Rotation}}} = 0.075$, $p = 0.790$). Log-transforming resolved normality issues ($W_{\text{\textsc{Scaling}}} = 0.989$, $p = 0.760$; $W_{\text{\textsc{Rotation}}} = 0.974$, $p = 0.144$). Mauchly's tests confirmed sphericity for Approach ($W_{\text{\textsc{Scaling}}} = 0.950$, $p = 0.772$; $W_{\text{\textsc{Rotation}}} = 0.584$, $p = 0.068$) and Approach \(\times\) Distance ($W_{\text{\textsc{Scaling}}} = 0.971$, $p = 0.862$; $W_{\text{\textsc{Rotation}}} = 0.997$, $p = 0.986$). All results presented below are based on log-transformed data.

The analysis revealed significant main effects of Approach for both \textsc{Scaling} ($F_{2, 22} = 9.00$, $p < 0.01$, $\eta^2_G = 0.128$, $1 - \beta = 1.00$) and \textsc{Rotation} ($F_{2, 22} = 27.27$, $p < 0.0001$, $\eta^2_G = 0.376$, $1 - \beta = 1.00$). Post hoc analyses indicated that both \textsc{Obj} and \textsc{NObj} significantly reduced Movement compared to \textsc{Hands} for \textsc{Scaling}\footnote{$\text{Est} = 0.352$, $\text{SE} = 0.080$, $t(11) = 4.38$, $p < 0.01$;\\ $\phantom{{}^{5}}\text{Est} = 0.295$, $\text{SE} = 0.098$, $t(11) = 3.02$, $p < 0.05$} and \textsc{Rotation}\footnote{$\text{Est} = 0.452$, $\text{SE} = 0.052$, $t(11) = 8.67$, $p < 0.0001$; \\ $\phantom{{}^{5}}\text{Est} = 0.616$, $\text{SE} = 0.103$, $t(11) = 6.00$, $p < 0.001$}. There was no significant difference between \textsc{Obj} and \textsc{NObj} for either \textsc{Scaling} ($\text{Est} = 0.057$, $\text{SE} = 0.088$, $t(11) = 0.65$, $p = 1.000$) or \textsc{Rotation} ($\text{Est} = 0.164$, $\text{SE} = 0.096$, $t(11) = 1.72$, $p = 0.342$). 

There was no significant main effect of Distance for either \textsc{Scaling} ($F_{1, 11} = 3.32$, $p = 0.096$, $\eta^2_G = 0.024$) or \textsc{Rotation} ($F_{1, 11} = 2.63$, $p = 0.133$, $\eta^2_G = 0.018$), and no significant interaction effect between Approach and Distance for \textsc{Scaling} ($F_{2, 22} = 2.51$, $p = 0.105$, $\eta^2_G = 0.034$) or \textsc{Rotation} ($F_{2, 22} = 0.16$, $p = 0.857$, $\eta^2_G  = 0.002$). 

\mySubsub{Time}
We fitted two-way repeated measures ANOVAs for \textsc{Scaling} and \textsc{Rotation}. Shapiro-Wilk tests indicated non-normality for \textsc{Rotation} ($W = 0.945$, $p < 0.01$), but normality for \textsc{Scaling} ($W = 0.981$, $p = 0.342$). The Kolmogorov-Smirnov tests showed that \textsc{Rotation} Time followed a log-normal distribution ($D = 0.089$, $p = 0.584$). Log-transforming \textsc{Rotation} Time resolved normality issues ($W = 0.980$, $p = 0.301$). Mauchly's tests confirmed sphericity for Approach ($W_{\text{\textsc{Scaling}}} = 0.706$, $p = 0.175$; $W_{\text{\textsc{Rotation}}} = 0.982$, $p = 0.915$) and Approach \(\times\) Distance ($W_{\text{\textsc{Scaling}}} = 0.939$, $p = 0.729$; $W_{\text{\textsc{Rotation}}} = 0.977$, $p = 0.888$). For \textsc{Rotation}, the results presented below are based on log-transformed data, while \textsc{Scaling} used the original data.

The analysis revealed significant main effects for Distance, with \textsc{Near} having significantly shorter Time than \textsc{Far}, for \textsc{Scaling} (\(F_{1, 11} = 6.42\), \(p < 0.05\), \(\eta^2_G = 0.033\)) and for \textsc{Rotation} (\(F_{1, 11} = 7.11\), \(p < 0.05\), \(\eta^2_G = 0.059\)). 

The analysis did not show main effects of Approach on either \textsc{Scaling} (\(F_{2, 22} = 0.17\), \(p = 0.845\), \(\eta^2_G = 0.003\), $1 - \beta = 0.156$) or \textsc{Rotation} (\(F_{2, 22} = 0.14\), \(p = 0.874\), \(\eta^2_G = 0.004\), $1 - \beta = 0.132$). To determine if different Approaches are equivalent in terms of Time, we used pairwise equivalence tests with the Two One-Sided Tests (TOST) procedure (equivalence bounds of Cohen's $d_z = \pm0.5$, $\alpha = 0.05$) and null hypothesis significance testing (NHST) using t-tests. NHST confirmed no significant differences among the three Approaches for both \textsc{Scaling} and \textsc{Rotation}. However, TOST did not yield significant equivalence, likely due to the limited statistical power inherent in the small sample size. Details in \Cref{table-time-equ}.

For \textsc{Scaling}, there was a significant interaction effect between Distance and Approach (\(F_{2, 22} = 6.04\), \(p < 0.01\), \(\eta^2_G = 0.075\)). However, post hoc analyses revealed no significant differences (\Cref{table-time-post-hoc}). For \textsc{Rotation}, there was no significant interaction effect between Distance and Approach ($F_{2, 22} = 1.94, p = 0.167, \eta^2_G = 0.026$). 

\vspace{1em}

\noindent\textit{\sffamily Summary.\,} 
For both \textsc{Scaling} and \textsc{Rotation} Tasks, \textsc{Obj} and \textsc{NObj} showed significantly lower \emph{Error} and \emph{Movement} than \textsc{Hands}, with no significant differences in \emph{Time} (though statistical equivalence was unconfirmed). In \textsc{Scaling}, \textsc{Obj} showed significantly lower \emph{Error} than \textsc{NObj}, while other comparisons between \textsc{Obj} and \textsc{NObj} showed no significant differences.

\vspace{-1.5em}
\begin{table}[H]
\caption{Raw and log-transformed means, read outward from the Metric column: the left side details Distance (Dist.), while the right details Approach (Appr.) and their interaction.}
    \Description{A dense, split-axis statistical table presenting raw and log-transformed means for three experimental metrics: Error, Movement, and Time, each further divided into Scaling and Rotation Tasks. The table is structurally centered around a central ``Metric'' column. To the left of this center axis, it details the main effect of Distance (Near and Far) alongside their corresponding raw and log means. To the right of the center axis, it details the main effect of Approach (Hands, Obj, NObj) and their means, sequentially followed by a further breakdown of the interaction effect between Approach and Distance (Near and Far) with their respective means.}
    \label{tab:means-table-all}
    \centering
    \resizebox{\columnwidth}{!}{%
    \setlength{\tabcolsep}{2.2pt} 
    \renewcommand{\arraystretch}{1.06}

    \begin{tabular}{cc c c c cc c cc} 
    \toprule
    \multicolumn{2}{c}{\textbf{Mean}} & \multirow{2}{*}{\raisebox{-0.35ex}[0pt][0pt]{\textbf{Dist.}}} & \multirow{2}{*}{\raisebox{-0.35ex}[0pt][0pt]{\textbf{Metric}}} & \multirow{2}{*}{\raisebox{-0.35ex}[0pt][0pt]{\textbf{Appr.}}} & \multicolumn{2}{c}{\textbf{Mean}} & \multirow{2}{*}{\raisebox{-0.6ex}[0pt][0pt]{\textbf{\begin{tabular}[c]{@{}c@{}}Appr.\\$\times$\\Dist.\end{tabular}}}} & \multicolumn{2}{c}{\textbf{Mean}} \\ 
    \cmidrule(lr){1-2} \cmidrule(lr){6-7} \cmidrule(lr){9-10} 
    \textbf{Log} & \textbf{Raw} & & & & \textbf{Raw} & \textbf{Log} & & \textbf{Raw} & \textbf{Log} \\ 
    \midrule 
    
    \multirow{3}{*}{$-4.11$} & \multirow{3}{*}{2.21} & \multirow{3}{*}{Near} & \multirow{6}{*}{\textbf{\begin{tabular}[c]{@{}c@{}}Scaling\\Error\\(\%)\end{tabular}}} & \multirow{2}{*}{Hands} & \multirow{2}{*}{7.29} & \multirow{2}{*}{$-2.90$} & Near & 3.62 & $-3.52$ \\ 
     & & & & & & & Far & 10.96 & $-2.28$ \\ 
     & & & & \multirow{2}{*}{Obj} & \multirow{2}{*}{1.61} & \multirow{2}{*}{$-4.29$} & Near & 1.06 & $-4.63$ \\ 
    \multirow{3}{*}{$-3.30$} & \multirow{3}{*}{5.30} & \multirow{3}{*}{Far} & & & & & Far & 2.15 & $-3.96$ \\ 
     & & & & \multirow{2}{*}{NObj} & \multirow{2}{*}{2.37} & \multirow{2}{*}{$-3.92$} & Near & 1.95 & $-4.17$ \\ 
     & & & & & & & Far & 2.79 & $-3.66$ \\ 
     \addlinespace
     
    \multirow{3}{*}{0.073} & \multirow{3}{*}{1.22} & \multirow{3}{*}{Near} & \multirow{6}{*}{\textbf{\begin{tabular}[c]{@{}c@{}}Rotation\\Error\\($^\circ$)\end{tabular}}} & \multirow{2}{*}{Hands} & \multirow{2}{*}{1.94} & \multirow{2}{*}{0.502} & Near & 1.34 & 0.161 \\ 
     & & & & & & & Far & 2.54 & 0.844 \\ 
     & & & & \multirow{2}{*}{Obj} & \multirow{2}{*}{1.35} & \multirow{2}{*}{0.174} & Near & 1.14 & 0.006 \\ 
    \multirow{3}{*}{0.492} & \multirow{3}{*}{1.88} & \multirow{3}{*}{Far} & & & & & Far & 1.56 & 0.342 \\ 
     & & & & \multirow{2}{*}{NObj} & \multirow{2}{*}{1.35} & \multirow{2}{*}{0.171} & Near & 1.18 & 0.050 \\ 
     & & & & & & & Far & 1.52 & 0.291 \\ 
    \midrule

    \multirow{3}{*}{$-0.714$} & \multirow{3}{*}{0.534} & \multirow{3}{*}{Near} & \multirow{6}{*}{\textbf{\begin{tabular}[c]{@{}c@{}}Scaling\\Movement\\(m)\end{tabular}}} & \multirow{2}{*}{Hands} & \multirow{2}{*}{0.601} & \multirow{2}{*}{$-0.561$} & Near & 0.572 & $-0.588$ \\ 
     & & & & & & & Far & 0.631 & $-0.533$ \\ 
     & & & & \multirow{2}{*}{Obj} & \multirow{2}{*}{0.446} & \multirow{2}{*}{$-0.913$} & Near & 0.479 & $-0.854$ \\ 
    \multirow{3}{*}{$-0.839$} & \multirow{3}{*}{0.480} & \multirow{3}{*}{Far} & & & & & Far & 0.414 & $-0.971$ \\ 
     & & & & \multirow{2}{*}{NObj} & \multirow{2}{*}{0.474} & \multirow{2}{*}{$-0.856$} & Near & 0.551 & $-0.699$ \\ 
     & & & & & & & Far & 0.397 & $-1.013$ \\ 
     \addlinespace
     
    \multirow{3}{*}{$-0.797$} & \multirow{3}{*}{0.502} & \multirow{3}{*}{Near} & \multirow{6}{*}{\textbf{\begin{tabular}[c]{@{}c@{}}Rotation\\Movement\\(m)\end{tabular}}} & \multirow{2}{*}{Hands} & \multirow{2}{*}{0.652} & \multirow{2}{*}{$-0.487$} & Near & 0.703 & $-0.429$ \\ 
     & & & & & & & Far & 0.600 & $-0.544$ \\ 
     & & & & \multirow{2}{*}{Obj} & \multirow{2}{*}{0.400} & \multirow{2}{*}{$-0.939$} & Near & 0.411 & $-0.911$ \\ 
    \multirow{3}{*}{$-0.889$} & \multirow{3}{*}{0.444} & \multirow{3}{*}{Far} & & & & & Far & 0.389 & $-0.966$ \\ 
     & & & & \multirow{2}{*}{NObj} & \multirow{2}{*}{0.367} & \multirow{2}{*}{$-1.103$} & Near & 0.391 & $-1.051$ \\ 
     & & & & & & & Far & 0.343 & $-1.155$ \\ 
    \midrule

    \multirow{3}{*}{N/A} & \multirow{3}{*}{4.51} & \multirow{3}{*}{Near} & \multirow{6}{*}{\textbf{\begin{tabular}[c]{@{}c@{}}Scaling\\Time\\(s)\end{tabular}}} & \multirow{2}{*}{Hands} & \multirow{2}{*}{4.73} & \multirow{2}{*}{N/A} & Near & 4.06 & N/A \\ 
     & & & & & & & Far & 5.40 & N/A \\ 
     & & & & \multirow{2}{*}{Obj} & \multirow{2}{*}{4.65} & \multirow{2}{*}{N/A} & Near & 4.54 & N/A \\ 
    \multirow{3}{*}{N/A} & \multirow{3}{*}{4.94} & \multirow{3}{*}{Far} & & & & & Far & 4.75 & N/A \\ 
     & & & & \multirow{2}{*}{NObj} & \multirow{2}{*}{4.80} & \multirow{2}{*}{N/A} & Near & 4.93 & N/A \\ 
     & & & & & & & Far & 4.68 & N/A \\ 
     \addlinespace
     
    \multirow{3}{*}{1.42} & \multirow{3}{*}{4.24} & \multirow{3}{*}{Near} & \multirow{6}{*}{\textbf{\begin{tabular}[c]{@{}c@{}}Rotation\\Time\\(s)\end{tabular}}} & \multirow{2}{*}{Hands} & \multirow{2}{*}{4.44} & \multirow{2}{*}{1.47} & Near & 4.03 & 1.37 \\ 
     & & & & & & & Far & 4.85 & 1.56 \\ 
     & & & & \multirow{2}{*}{Obj} & \multirow{2}{*}{4.35} & \multirow{2}{*}{1.46} & Near & 4.27 & 1.43 \\ 
    \multirow{3}{*}{1.52} & \multirow{3}{*}{4.63} & \multirow{3}{*}{Far} & & & & & Far & 4.42 & 1.48 \\ 
     & & & & \multirow{2}{*}{NObj} & \multirow{2}{*}{4.52} & \multirow{2}{*}{1.49} & Near & 4.43 & 1.46 \\ 
     & & & & & & & Far & 4.61 & 1.51 \\ 
    \bottomrule
    \end{tabular}
    } 
\end{table}

\subsection{Results: User Experience}\label{sec:experience}
The NASA-TLX and UEQ-S analyses showed no significant differences among Approaches (details in \Cref{appendix-nasa}). Based on the median rank given by the participants for their preferences (\Cref{fig:preferences}), \textsc{Obj} received the highest rank, followed by \textsc{NObj}, with \textsc{Hands} receiving the lowest rank. The Friedman test was not significant but indicated a trend ($\chi^2(2) = 5.17$, $p = 0.076$, $W = 0.022$).

\vspace{0.5em}
\begin{center}   
\includegraphics[width=\linewidth]{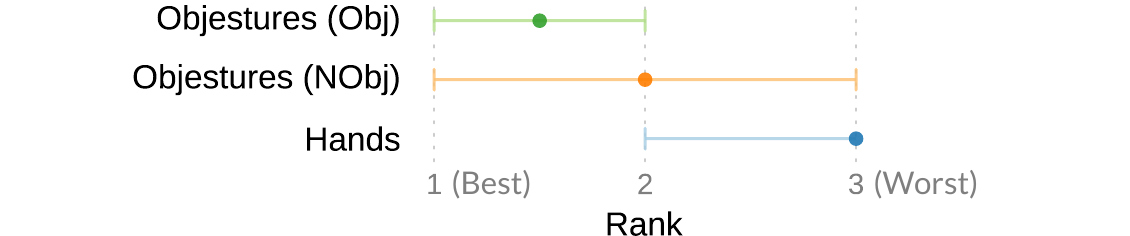}
\vspace{-2em}
    \captionof{figure}{Median preference ranks across three Approaches rated by all participants (lower values indicate higher preference). The error bars represent the interquartile range (IQR).}
    \Description{A horizontal dot plot displaying median preference ranks for three Approaches: Objestures (Obj), Objestures (NObj), and Hands. The x-axis is restricted to valid ranks of 1, 2, and 3. Each Approach is represented by a colored circular dot at its median rank with a horizontal error bar indicating the interquartile range (IQR). Both Obj and NObj achieve significantly higher preference (lower numerical rank) than Hands. Specifically, Obj is ranked best with a median of 1.5 and an IQR from 1 to 2. NObj follows with a median rank of 2, though it has the widest IQR, spanning from 1 to 3. Hands is ranked lowest with a median of 3 and an IQR from 2 to 3.}
    \label{fig:preferences}
\end{center}
Most participants found \textsc{Hands} direct and intuitive, especially for \textsc{Near} cubes (P0, P1, P3, P4, P5, P7, P9, P10). However, interacting with \textsc{Far} cubes was challenging due to the need to aim both rays at the cube (P1, P3, P5, P7, P11): ``I found it difficult to train the lasers on and then adjust properly because the motions were smaller and thus more sensitive''~(P3).

For both \textsc{Near} and \textsc{Far} cubes, the main issue with \textsc{Hands} was the lack of precision and stability due to the need for symmetric bimanual movements (P3, P4, P5, P7, P10). Especially when both hands unpinch, it is easy to move the cube slightly (P4, P8):  ``I didn't like having to use both hands for the same movement because it felt more destabilizing''~(P4).

\textsc{Obj} provided a sense of control, stability, and support due to the presence of physical objects (P1, P4, P5, P7, P8, P10, P11), and was perceived as innovative and fun (P0, P7, P9). The use of real-world objects provided a physical anchor, enabling peripheral awareness and exteroception (P10, P11): ``It's much more stable to control using a [physical] object. I don't need to look at it because I can feel where it is in reality''~(P10). 

The form of the objects offered cues that made interactions more intuitive, reminding participants of familiar controls (P4, P7, P8): ``the sliding motion [for \textsc{Scaling}] feels similar to using my computer's Touch Bar~\cite{apple_touchbar_2016}''~(P7). Flexibility in placing controllers anywhere was another advantage (P1). Participants also appreciated the asymmetrical movements (P2, P3, P4): ``although it feels nice to grab a cube with both hands, I think it is easier to manipulate with one hand grabbing and the other making the adjustment''~(P4).

However, there were limitations due to the physical properties of the objects, such as limited range of motion (P7, P11): ``when I start in the middle and the book just isn't long enough''~(P11), and potential danger from occupying physical space: ``I might accidentally knock over the coffee cup and spill the coffee''~(P5).

When physical objects were absent, \textsc{NObj} provided flexibility (P2, P3, P8, P9), ease of use (P3, P4, P6, P7, P11), and the translucent interaction zones were intuitive (P5, P10). However, the lack of physical objects introduced issues such as difficulty with depth perception and uncertainty about controller activation (P1, P10). This also led to fatigue and scattered attention (P3, P8, P10): ``more of my attention was on controlling the motion of my hands, as they didn’t have anything to rest on anymore''~(P3).

Therefore, some participants expressed a need for haptic feedback (P0, P6, P8, P9, P11). For example, having a book for sliding was preferred over not having one (P0, P6). 

While many participants found using everyday objects as controllers novel, those with little or no XR experience found purely virtual interactions more engaging (P4, P7, P11), though they appreciated the practicality of physical objects: ``It was cooler to do it without objects because you are really doing it in the air \ldots [but] if you actually want to use it in your daily life, objects might make it more convenient''~(P4).

\section{Case Studies}\label{sec:study2}
While the user study validated the basic usability of \oj{}, this section investigates its expressive potential. Users explored three applications we designed (\Cref{fig:teaser-full}A--C) that collectively cover all five interaction types (\Cref{fig:designSpace}). The participants and apparatus remained the same as in \Cref{sec:participants_apparatus}, with the addition of a cup lid and a stuffed animal to the setup. 

\begin{figure*}
\centering
\includegraphics[width=\textwidth]{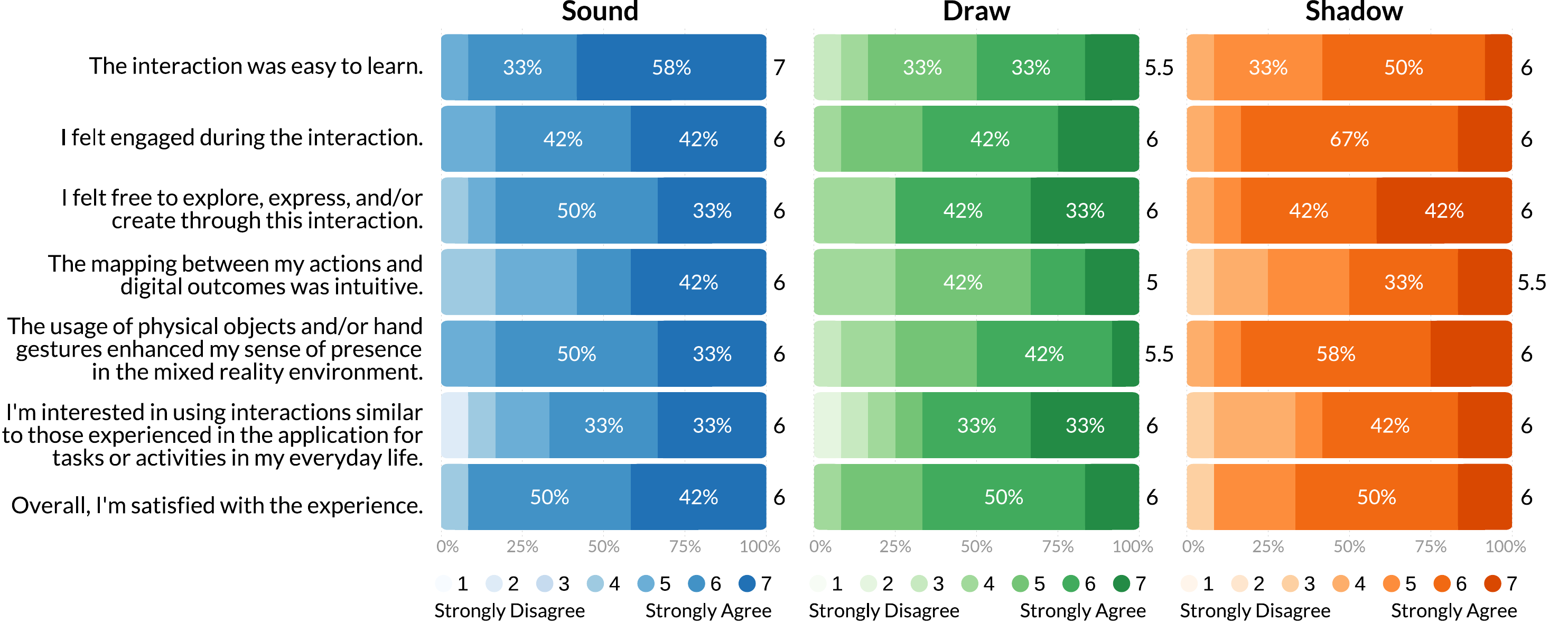}
\caption{Stacked bar plots showing user responses across three examples (\Cref{fig:teaser-full}A--C) on seven Likert items, color-coded from strongly disagree to strongly agree, with median values displayed to the right of the bars. Percentages for segments $\geq$~4 users (33~\%) are indicated for clarity. All applications had median ratings $\geq$~5.5 on all items.}
\Description{A data visualization consisting of three parallel columns of stacked horizontal bar charts, grouped by application and color-coded: ``Sound'' in blues, ``Draw'' in greens, and ``Shadow'' in oranges. Seven survey statements form the row labels on the left. Visually, the data is heavily skewed to the right across all three columns; the horizontal bars are dominated by the darkest shades representing positive responses (scores 5, 6, and 7), while the lightest shades representing disagreement are almost entirely absent. The large, dark blocks within each bar contain embedded percentage text (such as 33\%, 42\%, 50\%, and 58\%), and a single numeric median score aligns neatly to the right-hand edge of every column.}
\label{fig:study2-questionnaire}
\end{figure*}

To mitigate order effects, the three applications were fully counterbalanced. For each, participants received an explanation and explored freely without assigned tasks. Once satisfied, they completed a 7-item Likert questionnaire (\Cref{fig:study2-questionnaire}, left) and provided feedback on what they liked and disliked. Each application supported both left- and right-hand use. The mean total duration for completing all three applications was 38~min 23~s (SD = 9~min 32~s).

We used Friedman tests to analyze the differences in user responses across three applications on Likert scale items. Significant differences were found for the ease of learning ($\chi^2(2) = 8.60$, $p = 0.014$, Kendall's $W = 0.303$). Post-hoc analysis using the Wilcoxon signed-rank test with Bonferroni correction revealed that, compared to Shadow (median = 6, IQR = 5--6), Sound (median = 7, IQR = 6--7) was significantly easier to learn ($W = 0$, $p < 0.05$, $r = 0.796$). No significant differences were observed for the other items (see \Cref{table:study2-likert}).

\subsection{Sound}\label{sec:sound}
Street performers often create music using everyday objects, such as glass cups, buckets, or railings, but this requires finding items with the right tones or tuning them by hand. Imagine turning any object---even those that do not typically produce sound---into a musical instrument. This application enables this through a Nonlinear EOI and a Linear MAI: the user controls dynamics by squeezing a stuffed animal, and adjusts pitch by sliding their index finger along the edge of a hardcover book. Seven pitches appear as equally spaced spherical markers along this edge, defaulting to a violin sound in the C Major scale ($C_5$ to $B_5$). Thus, by sliding their finger near specific markers while squeezing the stuffed animal, users can play different pitches with expressive dynamics (\Cref{fig:teaser-full}, A1--A3).

Most participants enjoyed using both hands to interact with the physical objects (P0, P3, P5, P7, P10), finding the sound feedback satisfying and interactive (P6, P7). As P3 remarked, ``It was easy to understand \ldots adjusting the pitch was especially straightforward and easy to do.''

However, choosing the appropriate object appeared crucial. While the stuffed duck was generally easy to squeeze, some struggled to maintain consistent dynamics (P3, P5, P8, P10) or found it uncomfortable to grip with perspiring hands (P5). Beyond functional ergonomics, objects can also evoke emotional responses; for instance, some felt the duck was simply too cute to squeeze (P7, P11). 

Participants also envisioned alternative ways to interact with the objects, such as using multiple fingers to trigger simultaneous pitches (P7, P9), expecting continuous rather than discrete pitch changes while sliding (P2, P8, P11), or changing the duck's spatial position to alter the melody or tone (P7). Finally, given the application's playful nature, participants with musical backgrounds found the implementation somewhat oversimplified (P6, P7).

\subsection{Draw}\label{sec:draw}
Spatial drawing is an engaging experience for artists and non-artists alike~\cite{tan_wieldingcanvas_2024}, particularly in MR, where artwork can interact with real-world objects (e.g., drawing virtual liquid that appears to spray from a physical bottle). This application employs a Free MAI and a Rotational EOI: the user's dominant index finger acts as a pen, leaving a spatial trace when moving while the thumb and middle finger are pinched together. Concurrently, inspired by prior work~\cite{aslan_pen_2018, rodriguez_artists_2024} and products such as the Surface Dial~\cite{microsoft_dial}, the non-dominant hand rotates a coffee cup to dynamically adjust the brush size, yielding smooth, variable-width strokes. Holding a ``stop'' gesture (palm facing forward) with the drawing hand for 3~s clears the drawing.

Overall, participants found it to be an enjoyable and enabling experience (P1, P3, P4, P5, P7, P8, P11). The bimanual interaction felt natural and intuitive (P1, P2, P5, P9, P11)---especially for the drawing hand, which felt ``like grabbing a pen'' and ``using [their] hand as in real-life writing or drawing''~(P4, P10). The presence of physical objects greatly supported interaction (P9, P10). 

However, participants also identified areas for physical and functional improvement. Physically, some desired heavier objects for better stability (P10) and wished for a tangible prop---``something pointy''~(P8)---to hold in the drawing hand. Functionally, participants proposed alternative EOI mappings, such as adapting the Nonlinear squeezing from the Sound application (P8) or the Linear sliding from the \textsc{NObj} condition (P6). They also requested features like a wiping gesture to undo strokes (P7, P8, P11), color palettes (P0, P7, P8, P9, P11), and varied brush shapes (P11). They further envisioned additional bimanual synergies, such as keeping the drawing hand stationary while rotating the cup to draw perfect circles, or directly manipulating the drawing with both hands (P8).

Spatial drawing has limitations, such as depth issues~\cite{arora_experimental_2017} (P0, P5, P8), but participants saw its creative potential. For instance, P11 noted that the variable stroke width is ``really helpful for calligraphy'' and that 3D drawing on physical objects---like ``decorating my cup''---feels quite natural. Ultimately, as P7 suggested, such interactions could have a revolutionary impact on artists once mastered.

\subsection{Shadow}\label{sec:shadow}
In architecture, the relationship between light and shadow is crucial for ensuring proper sunlight distribution~\cite{mende_designing_2000}. To explore this interplay, this application employs a Free EOI and a Free MAI. Two coffee cup lids on the table serve as bases for a tall and a short virtual building. When a user's hand hovers over a lid and remains still for 3~s, the system anchors a building to it, allowing the user to reposition the building by moving the lid. Holding the lid steady for another 3~s locks the building in place. Concurrently, making a fist with the other hand transforms it into a virtual sun. By moving this hand freely in space, users can observe real-time changes in the shadows cast by the buildings. To ensure these shadows interact realistically with the physical furniture and surfaces, we pre-scanned the experimental space using the headset.

Overall, participants found the interaction satisfying, particularly praising the realism of the shadows (P3, P5, P7, P8, P11). They appreciated the bimanual coordination (P1, P2, P4) and the tangible presence of the physical objects (P6, P8, P9). Holding a physical lid provided a sense of ``psychological trust''~(P9); as P6 noted, ``it gave the buildings something solid, instead of just touching the air. It also helped to move the buildings around more easily.''

Regarding the EOI with lids, weight preferences varied: some preferred lighter objects for effortless movement (P4), while others favored heavier ones to prevent accidental shifts (P6). Interestingly, users occasionally tested the boundaries between the real and virtual worlds, such as attempting to stack buildings---an action that caused the physical lids to fall (P6, P11). Functionally, participants desired additional control, such as the ability to rotate the buildings (P0, P7, P8, P11). For the MAI, participants highlighted the importance of avoiding easily mis-triggered hand poses; for instance, the fist gesture was prone to false activations because users naturally rest their hands in this position on the table (P6, P7, P8). The visual feedback also induced psychological impacts, with P11 reporting a perceived sensation of heat in their ``sun'' hand. 

Regarding bimanual interaction, while the system strictly assigned one hand to the lids and the other to the sun, participants expressed a desire for role flexibility, such as enabling either hand to manipulate the lids (P2). Finally, the 3~s dwell time required to activate the lids forced users into asynchronous movements (P1); participants suggested that reducing this threshold would afford more fluid interactions (P5, P9, P10).

\section{Discussion}\label{sec:discussion}
Our exploratory user study and case studies highlight the promise of \oj{}, as it supports basic 3D tasks (rotation and scaling) with performance (error, movement, time) and user experience (perceived workload, subjective ratings, preference rankings) comparable to the status quo (the headset's native freehand manipulation), while enabling expressive and engaging applications.

\subsubsection*{Bimanual Synergy and Multimodal Flexibility}
\begin{figure*}
\centering
\includegraphics[width=\textwidth]{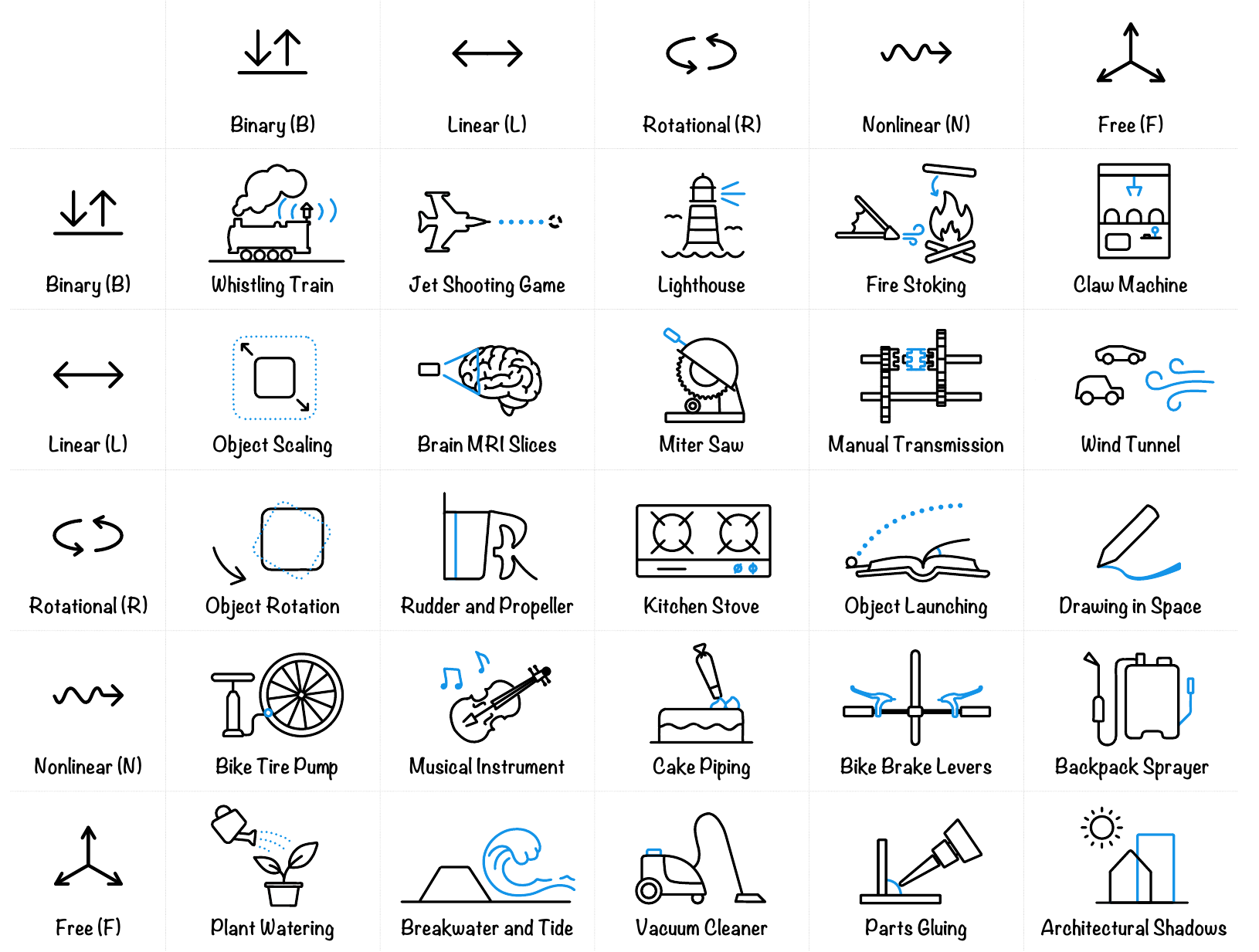}
\caption{Example applications that could be enabled by \oj{}, where each hand performs one of five interaction types (\Cref{fig:designSpace}) with either EOIs or MAIs. Diagonal cells depict symmetric interactions (same type in both hands). Off-diagonal cells show asymmetric ones. Detailed descriptions with five additional symmetric examples are in \Cref{sec:app-remaining}.}
\Description{A five-by-five grid matrix illustrating a design space of bimanual interaction combinations. The matrix rows and columns are both labeled with five interaction types: Binary (B), Linear (L), Rotational (R), Nonlinear (N), and Free (F), accompanied by motion-based icons. The resulting 25 intersecting cells each contain a line-art icon paired with a descriptive label, representing a specific application scenario that combines the corresponding row and column interaction types. For instance, the Binary-Binary intersection features a ``Whistling Train,'' the Linear-Rotational intersection shows a ``Miter Saw,'' and the Free-Free intersection depicts ``Architectural Shadows.'' Visually, the icons consistently use black lines to delineate the main physical objects and light blue accents to highlight the specific moving parts or interactive elements, such as the lever on the bike brakes or the trajectory of a drawn line.}
\label{fig:designSpace-potential}
\end{figure*}

Instead of piecemeal applications of EOIs and MAIs (\Cref{sec:intro}), \oj{} provides combinatorial building blocks that yield synergistic benefits: for instance, rather than merely acting as ``Hand + Hand,'' \textsc{NObj} delegates one hand for precision and the other for selecting, which yielded lower error and required less arm movement. Participants also found this delegation more stable than using hands symmetrically (\Cref{sec:experience}). This suggests that designers could make broader use of such delegation to further leverage the hands' symmetric and asymmetric potential. To this end, \Cref{fig:designSpace-potential} illustrates how \oj{} could enrich interactions across diverse contexts.

Our results further nuance the suggested trade-off between EOIs and MAIs~\cite{gong_affordance-based_2023}. Except for scaling error, we found no significant differences in performance or overall preference between \textsc{NObj} and \textsc{Obj}. However, individual variability exists in preferences for MAIs and EOIs due to perceived novelty (P4) or psychological trust (P9). Similar variability appeared across objects and gestures: participants held opposing views on the comfort of the stuffed animal (P7 vs.~P5) and the naturalness of drawing gestures (P1 vs.~P8). Hence, future work involving EOIs, MAIs, and other modalities should support flexible switching to accommodate varying preferences and contexts. For instance, a user might rotate an apple on a table to adjust speaker volume; if uncomfortable, they should be able to switch to a cup or use mid-air gestures instead.

\subsubsection*{Physical Anchoring and Semantic Congruence}
Compared to native freehand manipulations (\textsc{Hands}), utilizing physical objects (\textsc{Obj}) improves accuracy and reduces arm movement, mitigating pervasive MAI challenges such as depth uncertainty~\cite{arora_experimental_2017} and ``Gorilla Arm'' fatigue~\cite{hincapie-ramos_consumed_2014}.

Given EOIs' rarity in commercial products, we anticipated participants would praise their novelty. Interestingly, XR novices initially found purely virtual interactions (\textsc{NObj}, \textsc{Hands}) more exciting (\Cref{sec:experience}). This ``Better Because It's New'' effect~\cite{rutten_better_2020} reveals a bidirectional dynamic: XR veterans seek physical reconnection to counter digital fatigue, while novices favor virtual escapism from physical constraints. However, as MR inevitably blurs the digital--physical boundary~\cite{lyu_unbounded_2026}, the novelty of this escapism will fade. Rather than designing for short-term virtual gratification, practitioners should maintain a tangible presence. Everyday objects offer enduring value beyond mere input devices, acting as anchors for proprioceptive return (P10) and psychological trust (P9).

Crucially, leveraging these anchors requires navigating physical--digital semantic dissonance: when an object's assigned function misaligns with its real-world meaning, cognitive friction ensues~\cite{cooper_inmates_1999}. For instance, a stuffed duck felt too cute to squeeze (P7, P11), and rotating a coffee cup provoked anxiety about spilling (P5). \oj{} considers motion patterns of ``what can be done,'' but effective design must ultimately ask ``what \textit{should} be done,'' making semantic congruence paramount.

\subsubsection*{Limitations and Future Work}
Our results are based on an exploratory study and case studies with a small sample size ($N=12$). Although similar scales appear in prior XR work~\cite{zhu_bishare_2020, qian_fast-forward_2024, lee_gazepointar_2024}, we acknowledge this as a limitation; larger-scale evaluations are essential to further validate the findings. Beyond sample size, specific design choices likely shaped our outcomes. For example, in the exploratory study, the dominant hand was restricted to pinching (Binary) for simplicity. Given its capacity for fine-grained tasks~\cite{guiard_asymmetric_1987,buxton_study_1986}, future studies could also include translational movements (Free).

While prior work has noted hand tremor in VR~\cite{song_hotgestures_2023}---an issue our participants also reported---this may stem from tracking noise. Unlike VR, where physical hands are occluded, MR reveals that virtual hand overlays sometimes appear unstable relative to the physical hands. Furthermore, inferring object motion from the hand misses natural finger-only manipulations, such as rotating a cup by flicking the fingers while the hand remains still. Consequently, more accurate and direct methods (e.g., optical tracking) would better support these behaviors and may yield different results.

Moving forward, interviews~\cite{lyu_unbounded_2026} or workshops~\cite{jeon_sprayable_2025} with designers could investigate how the design space supports ideation and elicit unforeseen dimensions. For instance, the space could be extended to encompass motion phases~\cite{sharma_graspui_2024}, fine finger movements~\cite{sharma_solofinger_2021,joshi_transferable_2023,lee_grab-n-go_2025}, or other body parts (e.g., face~\cite{lee_designing_2018}, feet~\cite{wan_exploration_2024}, arms~\cite{lin_pub_2011}, ears~\cite{chen_exploring_2020}, mouth~\cite{xu_clench_2019})---offering accessible alternatives when certain modalities are constrained~\cite{kwon_accesslens_2024}. Ultimately, while \oj{} is not exhaustive, it provides a lens to highlight possibilities and invite further exploration. 

\section{Conclusion}
This paper addresses the lack of generalized guidance for integrating everyday object-based interactions and mid-air gesture interactions. By articulating scattered prior practices through \oj{}---five interaction types that form a design space---we provide designers and developers with a structured way to reason about the complementary properties of these two modalities. Through an exploratory user study and case studies using a prototype implementation, we show that \oj{} can support basic 3D tasks and enable expressive applications with positive user experiences. We further present 30 example applications to illustrate its potential. Together, this work serves as a building block for designing interactions that deepen the connection between the human, the physical world, and the digital experience.

\begin{acks}
The authors thank the Associate Chairs and the anonymous reviewers for their insightful comments and guidance, and members of the Intelligent Interactive Systems Group for valuable discussions. The first author was supported by the Trinity-Henry Barlow Scholarship and the Cambridge Trust International Scholarship.
\end{acks}

\bibliographystyle{ACM-Reference-Format}
\bibliography{objestures,additional}

\appendix
\counterwithin{table}{section}
\counterwithin{figure}{section}
\begingroup
\footnotesize

\captionsetup{
  font=footnotesize,
  labelfont={bf,footnotesize}
}
\myAppSection{Implementation Details}{appen:implementation-details}
Since the prototype does not rely on prior knowledge of objects, setup using Binary MAIs (e.g., thumbs-up) is required for EOIs across all types (\Cref{fig:setup}). Since MAIs and EOIs share the same interaction logic, we primarily discuss the implementation using examples based on EOIs.
 
\begin{center}   
    \includegraphics[width=\linewidth]{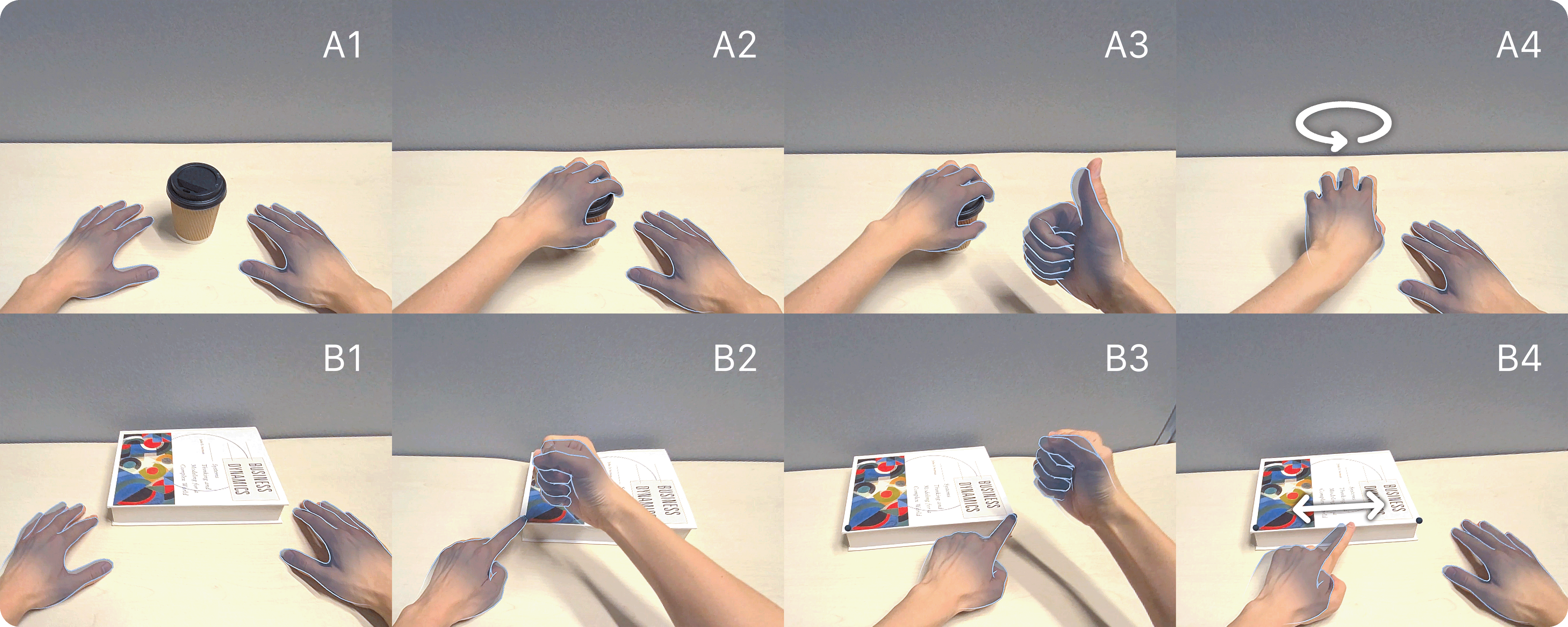}
    \captionof{figure}{The prototype we developed used Quest~3 hand tracking alone to infer movement for all five types. For example, cup rotation can be inferred from the middle-finger knuckle's rotation (A1--4), and sliding can be inferred from the index-finger position (B1--4). Here we show the setup: for Rotational with a cup (A1), place one hand on the cup (A2) and give a thumbs-up gesture with the other hand (A3); the cup's rotation is then tracked from hand movement (A4). For Linear along a book edge (B1), place an index finger at one end and make a hammer gesture (a downward striking motion with a closed fist) with the other hand (B2); repeat at the opposite end (B3). The sliding motion is then tracked (B4).}   
    \Description{A grid of eight first-person perspective MR captures arranged in two rows of four (A1--A4 and B1--B4), demonstrating a gestural setup sequence. The top row shows the user grasping a physical paper cup with their left hand, giving a thumbs-up gesture with their right hand, and subsequently rotating their left hand; this final rotation is annotated with a white, curved arrow. The bottom row shows the user touching the left edge of a physical book with their left index finger, making a downward fist gesture with their right hand, touching the right edge of the book while repeating the downward fist gesture, and finally sliding their left finger along the book's edge, which is annotated with a straight, white double-ended arrow.}
    \label{fig:setup}
\end{center}

\myAppSubSection{Binary}{appen:binary}
Whether dealing with EOIs or MAIs, we need to recognize two states: active and non-active. For MAIs, we use Oculus Integration's built-in Hand Pose Detection\footnote{\url{https://developer.oculus.com/documentation/unity/unity-isdk-hand-pose-detection/}}, which recognizes common gestures such as thumbs-up and also supports custom gestures. For EOIs, we use hand joint positions to define spatial interaction states. For example, when performing an EOI with a table bell, the system must register the spatial coordinates of the active (button fully pressed) and non-active (button released) states. During setup, the user places a designated tracking point (typically the dominant index fingertip) at the active position and performs a thumbs-up with the non-dominant hand to log the coordinate. This thumbs-up confirmation is then repeated for the non-active position. To account for hand-tracking jitter, we apply a 1~cm spatial threshold: the action triggers when the tracking point enters this radius around the active coordinate. For more complex interactions, active states can be customized based on specific spatial characteristics.

\myAppSubSection{Linear}{sys:linear}
We map the user's hand position within a defined movement range to a normalized output value. To configure a Linear interaction (e.g., sliding along a book edge), the user establishes the interaction axis by marking its two endpoints, $\mathbf{P}_1$ and $\mathbf{P}_2$, using a specific confirmation gesture (\Cref{fig:setup}, B1--B4). Once calibrated, either physical edges (EOIs) or translucent virtual zones (MAIs) guide the user's sliding movement. The raw output value $s$ is calculated by projecting the finger's current position onto the line segment vector $\mathbf{P}_2 - \mathbf{P}_1$, normalizing the result, and clamping it to a $[0, 1]$ range. To minimize hand-tracking jitter while maintaining responsiveness, we apply an Adaptive Exponential Moving Average (AEMA)~\cite{hunter_exponentially_1986} to filter $s$:
$$s = \alpha s + (1 - \alpha) s_{\text{prev}}, \quad \text{where} \ \alpha = \text{Lerp}(\alpha_{\text{min}}, \alpha_{\text{max}}, k \frac{\Delta s}{\Delta t})$$
The smoothing factor $\alpha$ is dynamically adjusted using linear interpolation (Lerp) between $\alpha_{\text{min}}$ (0) and $\alpha_{\text{max}}$ (1). This adjustment is driven by the rate of change $\frac{\Delta s}{\Delta t}$ (with $\Delta t = 1/60$~s at 60 FPS) and a sensitivity scaling factor $k$. Here, $s_{\text{prev}}$ refers to the filtered value from the previous time step. This adaptive approach allows the system to respond instantly to rapid movements (higher $\alpha$) while heavily smoothing slow, deliberate adjustments (lower $\alpha$). The sliding interaction activates only when the finger is within a 3~cm spatial threshold of the defined line; otherwise, the output value remains unchanged. To overcome physical length constraints, the system supports an optional ``infinite continuous sliding'' feature: if the finger reaches the physical boundary, exits the tracking zone, and re-enters at the middle, the output value resumes adjusting relative to the new contact point, enabling unbounded input along a finite physical edge.

\myAppSubSection{Rotational}{appen:lineary}
We estimate rotational output by tracking the relative angular displacement of a designated hand joint. For example, when rotating a coffee cup like a dial (EOI) using a downward grip~\cite{schlesinger_mechanische_1919, mackenzie_grasping_1994, aslan_pen_2018, dim_designing_2016}, the system tracks a central point (e.g., the middle finger's knuckle) and calculates its rotation around the vertical axis. During setup, the user logs this joint's initial position and orientation via a confirmation gesture (\Cref{fig:setup}, A1--A4). As the user rotates the object, the angular difference from this initial reference determines the rotational output. To accommodate the hand's limited anatomical range, the system supports a clutching mechanism: if the hand moves away from the initial tracking point, the system records the rotation at the point of departure. Upon re-entry, it resets the orientation reference and calculates new angular changes relative to this updated position, accumulating them seamlessly to maintain continuous rotation. For MAIs, the same logic tracks mid-air joint rotation directly without a physical proxy. We apply the AEMA filter (\Cref{sys:linear}) to mitigate jitter. Furthermore, to enhance precision without sacrificing speed, we implement an adaptive dynamic scaling factor. Each frame's raw rotation delta ($\Delta \theta$) is scaled by a multiplier $v$, which is linearly interpolated between customizable limits $v_{\text{min}}$ and $v_{\text{max}}$ based on the angular velocity:
$$ \theta \mathrel{+}= \Delta \theta \cdot v, \quad \text{where} \ v = \text{Lerp}(v_{\text{min}}, v_{\text{max}}, k \frac{\Delta \theta}{\Delta t}) $$
Here, $k$ is a sensitivity scaling factor, with $v_{\text{max}} > 1$ and $0 < v_{\text{min}} \leq 1$. Consequently, fast rotations are amplified ($v > 1$) for quick macroscopic adjustments, while slow rotations are attenuated ($v < 1$) to facilitate precise micro-tuning.

\myAppSubSection{Nonlinear}{appen:nonlinear}
For Nonlinear EOIs (e.g., squeezing a stuffed animal), we estimate manipulation intensity using a distance-based heuristic. The system uses the palm center as the origin and measures the distances from the five fingertips to this center.  During setup, the user rests their hand on the uncompressed object, and a confirmation gesture from the non-dominant hand logs the baseline sum of these distances ($D_{\text{baseline}}$). As the user squeezes the object, the system continuously updates the current distance sum ($D_{\text{current}}$). The normalized squeeze intensity is then calculated as $1 - D_{\text{current}} / D_{\text{baseline}}$, mapping the physical deformation to a continuous $[0, 1]$ output value. For MAIs, the same heuristic maps mid-air clenching directly to the output intensity.

\myAppSubSection{Free}{appen:free}
For Free EOIs (e.g., arranging coffee cup lids), the system tracks the 3D position of the user's hand. During setup, the user touches the physical object, and a confirmation gesture logs the distance between the contacting fingers (e.g., the index finger and thumb). The midpoint of these fingers approximates the object's spatial position, $\mathbf{P}_{\text{object}}$. To differentiate intentional manipulation from incidental movement, we implement a dwell-time selection mechanism.  When the fingers' midpoint enters a spatial tolerance zone around $\mathbf{P}_{\text{object}}$ and remains there for 3~s, the system assumes an intent to interact. As the hand moves, $\mathbf{P}_{\text{object}}$ continuously updates. If the user holds the object stationary for another 3~s, the system detaches it, ceasing updates to $\mathbf{P}_{\text{object}}$ until the hand re-enters the tolerance zone for a new dwell phase. For Free MAIs, the hand or finger is tracked directly.
\myAppSection[nospaceafter]{Supplementary User Study Results}{appen:results}

\myAppSubSection[nospaceafter]{Time}{appendix-time}
\vspace{-2em}
\begin{table}[H]
    \centering
    \footnotesize
    \setlength{\tabcolsep}{1.5pt} 
    \renewcommand{\arraystretch}{1} 
        \caption{Results of pairwise equivalence tests of Approach on Time for \textsc{Scaling} and \textsc{Rotation}}
        \Description{A statistical results table presenting pairwise equivalence test results. The first two columns list the pairwise comparisons (Hands vs. Obj, Hands vs. NObj, and NObj vs. Obj) and the Task (Rotation, Scaling). The remaining columns are divided into two major header blocks: ``TOST'' and ``NHST''. Under TOST, sub-columns report the t-statistic (df = 11), p-value, and 90\% confidence interval (CI). Under NHST, sub-columns report the t-statistic (df = 11), p-value, and 95\% CI. The cells contain the corresponding numerical results.}
    \label{table-time-equ}
    \begin{tabular}{ll ccc ccc}
    \toprule
    \multirow{2}{*}{\raisebox{-0.75ex}[0pt][0pt]{\textbf{Contrast}}} & \multirow{2}{*}{\raisebox{-0.75ex}[0pt][0pt]{\textbf{Task}}} & \multicolumn{3}{c}{\textbf{TOST}} & \multicolumn{3}{c}{\textbf{NHST}} \\ 
    \cmidrule(lr){3-5} \cmidrule(lr){6-8}
    & & $t(11)$ & $p$ & 90\% CI & $t(11)$ & $p$ & 95\% CI \\ 
    \midrule
    
    \multirow{2}{*}{Hands $-$ Obj} 
    & Scaling  & $-1.39$ & 0.097 & [$-0.349$, 0.515] & 0.35 & 0.736 & [$-0.446$, 0.613] \\ 
    & Rotation & $-1.59$ & 0.071 & [$-0.091$, 0.107] & 0.15 & 0.886 & [$-0.113$, 0.129] \\ 
    \addlinespace 
    
    \multirow{2}{*}{Hands $-$ NObj} 
    & Scaling  & 1.40 & 0.095 & [$-0.465$, 0.319] & $-0.33$ & 0.745 & [$-0.554$, 0.408] \\ 
    & Rotation & 1.35 & 0.102 & [$-0.114$, 0.074] & $-0.38$ & 0.710 & [$-0.135$, 0.095] \\ 
    \addlinespace
    
    \multirow{2}{*}{NObj $-$ Obj} 
    & Scaling  & $-1.26$ & 0.116 & [$-0.442$, 0.754] & 0.47 & 0.648 & [$-0.577$, 0.889] \\ 
    & Rotation & $-1.25$ & 0.118 & [$-0.077$, 0.133] & 0.48 & 0.642 & [$-0.101$, 0.157] \\ 
    \bottomrule
    \end{tabular}
\end{table}
\begin{table}[H]
\vspace{-2em}
    \centering
    \footnotesize
    \setlength{\tabcolsep}{6pt} 
    \renewcommand{\arraystretch}{1}
    \caption{Results of post hoc pairwise comparisons of Distance on Time for \textsc{Scaling}}
\Description{A six-column statistical table detailing post hoc comparison results. The columns are labeled Contrast, Distance, Est (Estimate), SE (Standard Error), t(11), and p (p-value). The rows are grouped into three primary Contrast pairs: Hands - NObj, Hands - Obj, and NObj - Obj. Each contrast pair is further split into two rows for ``Far'' and ``Near'' Distance conditions, followed by their corresponding statistical numerical values across the remaining four columns.}
\label{table-time-post-hoc}
    \begin{tabular}{ll cccc}
    \toprule
    \textbf{Contrast} & \textbf{Distance} & \textbf{Est} & \textbf{SE} & \textbf{$t(11)$} & \textbf{$p$} \\ 
    \midrule
    
    \multirow{2}{*}{Hands $-$ Obj}  
    & Near & $-0.483$ & 0.354 & $-1.37$ & 0.598 \\ 
    & Far  & 0.649  & 0.341 & 1.90  & 0.251 \\ 
    \addlinespace 
    
    \multirow{2}{*}{Hands $-$ NObj} 
    & Near & $-0.871$ & 0.376 & $-2.32$ & 0.122 \\ 
    & Far  & 0.725  & 0.282 & 2.57  & 0.078 \\ 
    \addlinespace
    
    \multirow{2}{*}{NObj $-$ Obj}   
    & Near & 0.388  & 0.465 & 0.83  & 1.000 \\ 
    & Far  & $-0.076$ & 0.299 & $-0.25$ & 1.000 \\ 
    \bottomrule
    \end{tabular}
\end{table}
\myAppSubSection{NASA-TLX and UEQ-S}{appendix-nasa}
We conducted a series of Friedman rank sum tests to compare differences in TLX scores across three Approaches for all six subscales (\Cref{fig:nasa-tlx}). Results showed a significant effect of Approach on Physical Demand ($\chi^2(2) = 7.68, p = 0.021, \text{Kendall's W} = 0.740$). However, post-hoc analysis using the Wilcoxon signed-rank test with Bonferroni correction found no significant differences (\textsc{Hands} vs. \textsc{Obj}: $W = 57$, $p = 0.104$; \textsc{Hands} vs. \textsc{NObj}: $W = 24$, $p = 1.000$; \textsc{Obj} vs. \textsc{NObj}: $W = 12.5$, $p = 0.118$). No significant effects were found for Effort ($\chi^2(2) = 1.59, p = 0.452, W = 0.778$), Frustration ($\chi^2(2) = 0.39, p = 0.823, W = 0.652$), Mental Demand ($\chi^2(2) = 4.22, p = 0.121, W = 0.707$), Performance ($\chi^2(2) = 5.90, p = 0.052, W = 0.833$), and Temporal Demand ($\chi^2(2) = 1.80, p = 0.407, W = 0.826$).

\begin{center}   
\vspace{0.3em}
    \includegraphics[width=\linewidth]{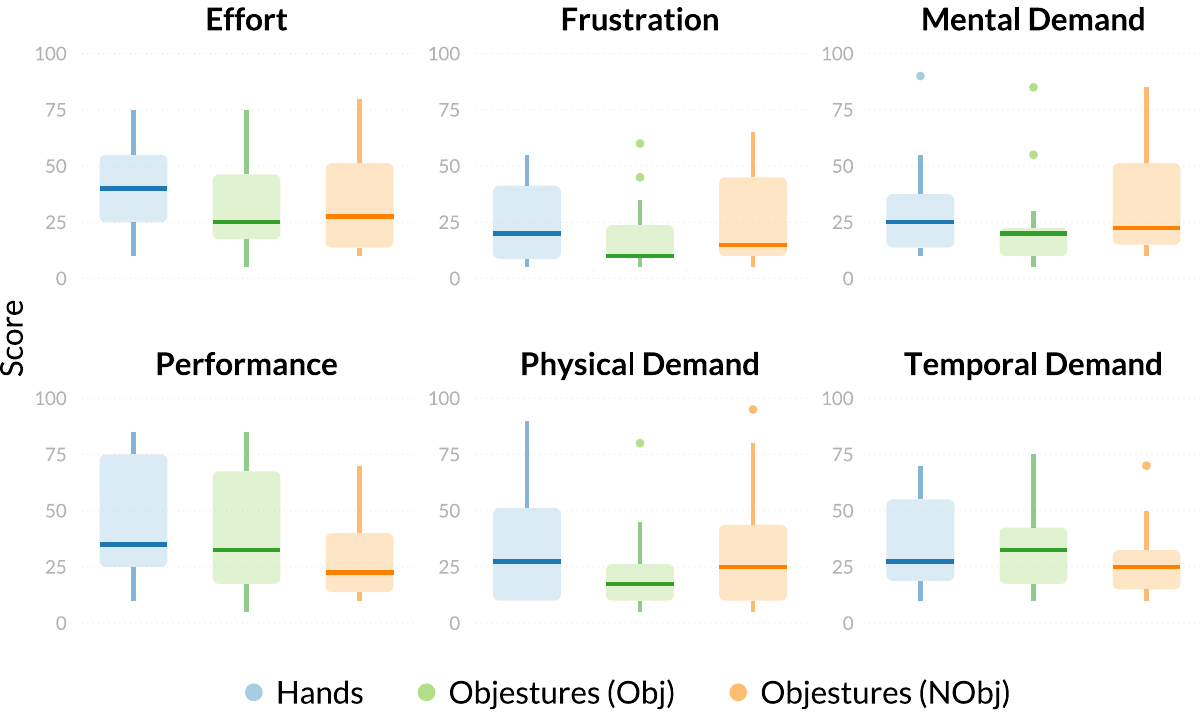}
    \vspace{-2.2em}
    \captionof{figure}{\footnotesize Box plots of NASA-TLX subscale scores across three Approaches. Lower scores indicate better performance for Performance, and lower demands or levels for other subscales. The boxes represent the interquartile range (IQR) with the median shown as a horizontal line inside each box. The whiskers extend to 1.5 times the IQR from the quartiles, and dots represent outliers beyond the whiskers.}
    \Description{A two-by-three grid of box plots displaying NASA-TLX scores across six subscales: Effort, Frustration, Mental Demand, Performance, Physical Demand, and Temporal Demand. Each chart compares three Approaches side-by-side, color-coded as Hands (blue), Objestures (Obj) (green), and Objestures (NObj) (orange). The y-axis for all charts ranges from 0 to 100. Visually, the distributions appear broadly similar across Approaches. Median scores generally rest in the lower half of the scale, roughly between 15 and 45. A few scattered outlier dots appear above the upper whiskers in the Frustration, Mental Demand, Physical Demand, and Temporal Demand charts, but the core interquartile ranges remain closely aligned within each metric.}
    \label{fig:nasa-tlx}
\end{center}
Each of the eight questions in the UEQ-S was evaluated using the Friedman test to determine significant differences across the scales. The test results indicated no significant differences for any individual question, with p-values ranging from 0.105 to 0.901 and Kendall's W values between 0.551 and 0.815. Although averaging ordinal data is not ideal, we followed the UEQ's approach for comparison with established benchmarks. The responses were aggregated into two scales: Pragmatic Quality (PQ) and Hedonic Quality (HQ). Based on data from 21,175 participants across 468 UEQ studies\footnote{\url{https://www.ueq-online.org/}}, the mean HQ scores for \textsc{Hands} (1.67), \textsc{Obj} (1.75), and \textsc{NObj} (1.85) surpass the Excellent benchmark (1.59). For PQ, only \textsc{Obj} (1.92) exceeds the Excellent benchmark (1.74), while the mean scores for \textsc{NObj} and \textsc{Hands} are 1.44 and 1.56, respectively.

\myAppSection[nospaceafter]{Additional Results for Case Studies}{appen:results-2}
\begin{table}[H]
\vspace{-2em}
\centering
\caption{Friedman tests results for user responses
across three applications on Likert scale items.}
\Description{A four-column statistical summary table. The left-most column, ``Measure'', lists seven subjective survey items, such as Ease of learning, Engagement, and Overall satisfaction. The three right-hand columns provide the corresponding statistical outputs for each measure: a Chi-square value with 2 degrees of freedom, a p-value, and Kendall's W coefficient. The table contains numerical data aligned across all rows.}
\label{table:study2-likert}
\begin{tabular}{lccc}
\toprule
\textbf{Measure} & \textbf{$\chi^2(2)$} & \textbf{$p$} & \textbf{$W$} \\
\midrule
Ease of learning & 8.60 & 0.014 & 0.303 \\
Engagement & 1.80 & 0.407 & 0.525 \\
Freedom to explore & 0.87 & 0.648 & 0.344 \\
Intuitiveness & 1.80 & 0.407 & 0.381 \\
Enhancement of presence & 3.07 & 0.215 & 0.344 \\
Interest in everyday use & 0.56 & 0.755 & 0.319 \\
Overall satisfaction & 4.19 & 0.123 & 0.512 \\
\bottomrule
\end{tabular}
\end{table}

\myAppSection{Example Applications}{sec:app-remaining}
Each paragraph title includes a reference in ``[ ]'', corresponding to a cell in \Cref{fig:designSpace-potential}. For instance, [LN] represents the Linear and Nonlinear interaction in the second row, fourth column. For each example, we provide both a recommended interaction method and an equivalent alternative (i.e., EOIs for MAIs and vice versa). For the five applications along the diagonal, since \Cref{fig:designSpace-potential} can only depict one example per cell, we use the format ``[XX-2]'' (e.g., [BB-2]) to indicate the second application not shown. 

In some examples, we also incorporate interactions not included in the initial design space (\Cref{fig:designSpace}), such as micro-gestures, as our subsequent studies and reflections suggested that the framework has broader applicability.


\paragraph{Whistling Train [BB]} A train runs on tracks controlled by two separate Binary interactions: one for movement and one for the whistle sound. For instance, one hand engages a latch: when engaged, the train moves forward; when disengaged, it stops, similar to an electrical switch. Alternatively, the hand can perform MAIs, such as a thumbs-up gesture to move the train forward or a stop gesture to halt it. The other hand can control the whistle sound by pressing a table bell, or by using a pinch gesture in mid-air.

\paragraph{Finger Whac-A-Mole [BB-2]}
Each hand taps the moles using Binary interactions. For example, both hands can hold a blister pack, with virtual moles appearing on the raised bumps (where medication is stored). The thumbs can press down on these bumps to hit the moles; when a mole is hit, it will appear in another position. If no physical object is available, the moles can appear arranged along the length of the index finger, allowing the thumb to tap different positions to hit them.

\paragraph{Jet Shooting Game [BL]} As shown in Figures~\ref{fig:teaser-full}~(D1--3), the jet's left and right movement is controlled by Linear interactions, while missile firing is controlled by Binary interactions. For example, one hand can slide along the edge of a desk to control the jet's lateral movement. If a physical object is unavailable, this hand can perform a left-right sliding gesture in mid-air instead. The other hand can press a table bell to control firing, or use the pinch gesture in mid-air.

\paragraph{Lighthouse [BR]} The lighthouse light's rotation is controlled by Rotational interactions, and its on/off state is controlled by Binary interactions. For instance, one hand can rotate an apple to control the light's angle, or perform a circular rotating gesture in mid-air. The other hand can perform a ``radiating light'' gesture (fingers spread outwards from a central point) to turn the light on, and retract the fingers to turn it off. Alternatively, pressing a table bell can achieve the same effect.

\paragraph{Fire Stoking [BN]} The intensity of blowing air to the flames is controlled by Nonlinear interactions, while adding wood to the fire is controlled by Binary interactions. For example, one hand squeezes a blower bulb, with the side of the mouth facing the flames, acting as a bellows (alternatively, this can be controlled by a mid-air squeezing gesture). Repeatedly blowing can make the flames grow stronger. As the wood gradually burns down, the other hand performs a downward flip gesture near the fire to add a piece of wood (alternatively, pressing a table bell can achieve the same effect).

\paragraph{Claw Machine [BF]}
The direction of the claw is controlled by Free interactions, while releasing the claw to grab objects is controlled by Binary interactions. For example, one hand shakes a water bottle, using the table as support, as if it were a joystick (alternatively, mid-air gestures can mimic holding and moving a joystick). Once the target is positioned correctly, the other hand presses a table bell to release the claw (alternatively, a mid-air gesture where all fingers curl inward to form a grabbing pose can release the claw; when the pose is not made, the claw does not release).

\paragraph{Object Scaling [LB]}
As in the \textsc{Scaling} Task in the user study, selecting an object is controlled by Binary interactions, while scaling its size is controlled by Linear interactions. For example, one hand slides on the edge of a hardcover book to adjust the object's size (alternatively, a finger can slide horizontally in mid-air as if adjusting an invisible slider). The other hand uses a pinch gesture to select the object to be manipulated (alternatively, this hand can hold a pen, approach the object, and click the button at the back of the pen to confirm selection).

\paragraph{Brain MRI Slices [LL]}
The presentation of brain slices in three directions---upper \& lower (Axial Plane), left \& right (Sagittal Plane), and front \& back (Coronal Plane)---is controlled by three Linear interactions. For example, using three mutually orthogonal edges of a table corner, each edge controls the slices on a specific plane. One hand can slide along the edge in the front and back direction to display the Coronal Plane cross-sections; the other hand can slide along the edge in the left and right direction to display the Sagittal Plane cross-sections. Alternatively, this sliding can be performed in mid-air with both hands, fingers straight and together, aligning each hand to a specific plane. For instance, one hand can align with the Axial Plane and move up and down to view slices at that position, while the other hand can align with the Sagittal Plane and move left and right. This is similar to~\cite{cordeil_embodied_2020}, but does not require dedicated tangible devices for moving along each axis.


\paragraph{Treadmill [LL-2]}
The treadmill's speed is controlled by a Linear interaction, as is the incline. For example, while running, sliding the left hand along the left handrail adjusts the treadmill's speed, and sliding the right hand along the right handrail adjusts the incline. If no physical handrails are available, these controls can also be managed by performing sliding hand motions in specific mid-air regions.

\paragraph{Miter Saw [LR]}
The movement of the wood along the fence is controlled by Linear interactions, while the saw's downward swinging motion to cut is controlled by Rotational interactions. For example, one hand slides along the edge formed by a box on a table, simulating pushing the wood along the fence (alternatively, a half-fist in mid-air mimicking gripping a handle and swinging down can be used to cut the wood). At the end of the box, the other hand controls a stapler, with the downward movement determining the cut depth (alternatively, a half-fist in mid-air mimicking holding a handle swinging down can be used to cut the wood). A paper cutter can also be used, with a piece of paper representing the wood, moved horizontally along the fence, while the other hand uses the cutter's handle to cut.

\paragraph{Manual Transmission [LN]}
Using a simple 2-Speed Sliding Mesh Transmission as an example, the input shaft (left part of the upper row of shafts and gears) speed is controlled by Nonlinear interactions, and the clutch (middle structure of the upper row) movement is controlled by Linear interactions. For instance, one hand can press the lever of a spray bottle to simulate the nonlinear feeling of pressing a gas pedal, with the degree of pressing affecting the input shaft's speed (alternatively, in mid-air, bending a finger on the same hand using the thumb can control the input shaft's speed). The other hand performs a mid-air gesture mimicking holding the shaft and sliding horizontally to move the clutch along the shaft. Sliding to the left engages the output shaft (right part of the upper row) with the same speed as the input shaft (high gear, suitable for high-speed driving). Sliding to the right engages the counter shaft below, resulting in lower speed but higher torque for the output shaft (low gear, suitable for starting). This can also be achieved by sliding along the edge of a rectangular tissue box. Adding more Linear controls can manage more gears, such as a three-speed transmission and a reverse gear~\cite{handy_jam_organization_spinning_1936}.

\paragraph{Wind Tunnel [LF]}
The airflow is visualized with lines to observe the drafting effects and aerodynamic interactions of vehicles, especially racing cars. The relative positions of the vehicles are controlled by Free interactions, while the wind intensity is controlled by Linear interactions. For example, one hand can arrange two objects on a table, such as two oranges, boxes, or toy car models (if objects are unavailable, virtual cars can be arranged in mid-air). The other hand can adjust the wind intensity by sliding the thumb along the index finger in mid-air (alternatively, this can be done by sliding along the edge of a book).

\paragraph{Object Rotation [RB]}
As in the \textsc{Rotation} Task in the user study, selecting an object is controlled by Binary interactions, and the rotation angle is controlled by Rotational interactions. For example, one hand can rotate a coffee cup to control the rotation of the object (alternatively, in mid-air, mimicking holding and rotating a coffee cup). The other hand pulls a zipper on a piece of clothing; the extent of the upward pull corresponds to the increase in propeller speed (alternatively, this can be achieved by sliding the thumb along the index finger in mid-air).

\paragraph{Rudder and Propeller [RL]}
The ship's rudder is controlled by Rotational interactions, and the propeller's speed is also controlled by Linear interactions. For example, rotating a cup on the table controls the rudder's angle (alternatively, in mid-air, the palm can mimic the shape of a rudder and rotate to control the angle). The other hand pulls a zipper on a piece of clothing; the degree of upward pull corresponds to the increase in propeller speed (alternatively, this can be achieved by sliding the thumb along the index finger in mid-air).

\paragraph{Kitchen Stove [RR]}
Two Rotational interactions control the heat levels of two burners. For example, each hand rotates a key in a keyhole, with the rotation angle representing the burner's heat level (alternatively, mid-air gestures mimicking the turning of control knobs can be used).

\paragraph{Water Faucets [RR-2]}
The cold and hot water taps are controlled by two separate Rotational interactions. For example, each hand rotates a coffee cup, with the degree of rotation representing the flow of cold and hot water, respectively. If no cups are available, rotating motions in mid-air can also control the water flow. The faucets and water flow can be either real or virtual.

\paragraph{Object Launching [RN]}
Launching a small ball into the air involves controlling the launch angle with Rotational interactions and the launch force with Nonlinear interactions. Similar to the game Angry Birds~\cite{angrybirds_official}, one hand performs a flick gesture, with the speed of the flick determining the launch force (alternatively, this hand can press down on a flexible ruler or bookmark held by a book, where the degree of press determines the rebound force). The other hand moves a book page, with the angle between the page and the horizontal plane determining the launch angle. When the object is unavailable, a mid-air gesture with a flat palm can be used, and the angle of the palm relative to the horizontal plane determines the launch angle.

\paragraph{Drawing in Space [RF]}
As in the Draw (\Cref{sec:draw}), the brush is controlled by Free interactions, and the brush size is controlled by Rotational interactions. For example, one hand makes an index finger pointing gesture, as if it were a pen, with the fingertip acting as the pen tip to draw in space (alternatively, a real pencil can be held to draw in space). The other hand rotates a coffee cup on the table, with the rotation degree controlling the brush size (alternatively, in mid-air, mimicking the gesture of holding and rotating a coffee cup).

\paragraph{Bike Tire Pump [NB]}
The motion of the pump handle is controlled by Nonlinear interactions, and the locking and unlocking of the pump head is controlled by Binary interactions. For example, one hand can push down on a sofa cushion, simulating the use of a pump (alternatively, this can be controlled by pressing the palm with the other four fingers in mid-air). Once the pump head is placed correctly on the valve, a pinch gesture on the lever above the pump head indicates locking (allowing air to flow in smoothly), while releasing the pinch indicates unlocking (preventing air flow). This can also be achieved using EOIs, such as pressing the push button on a pen: moving it near the pump head, pressing the button to lock, or releasing it to unlock.

\paragraph{Musical Instrument [NL]}
As in the Sound (\Cref{sec:sound}), the dynamics (relative changes in volume within a piece of music to convey emotions and artistic expression) are controlled by Nonlinear interactions, and the pitch is controlled by Linear interactions. For example, one hand can squeeze a stuffed animal, with the degree of squeezing corresponding to the dynamics (alternatively, a squeezing motion can be performed in mid-air). The other hand can slide along the edge of a book to control the pitch of the instrument (alternatively, mid-air gestures can be used, such as bending each of the five fingers, where each combination of bent fingers corresponds to different pitches).

\paragraph{Cake Piping [NR]}
The squeezing of the piping bag is controlled by Nonlinear interactions, while the rotation of the cake is controlled by Rotational interactions. For example, one hand can squeeze a stress ball above the cake, with the hand's position remaining stationary, and the degree of squeezing corresponding to the amount and speed of the cream being piped (alternatively, this can be done with a mid-air gesture mimicking the squeezing of a piping bag). The other hand can rotate a coffee cup to spin the cake (alternatively, in mid-air, a gesture mimicking the rotation of a coffee cup can be used). This allows for piping a circle of cream along the edge of the cake.

\paragraph{Bike Brake Levers [NN]}
The squeezing of the two brake levers is controlled by Nonlinear interactions. For example, each hand can grip the handle of a spray bottle, with the degree of squeezing corresponding to the braking force applied. Holding the spray bottle vertically simulates the feel of vertical brake levers (drop bar) on a road bike. If no suitable object is available, the braking force can also be controlled by making a squeezing gesture in mid-air to mimic squeezing brake levers.

\paragraph{CPR [NN-2]}
During CPR training, both hands perform Nonlinear interactions for control. Ideally, training should be done on a proper dummy body, but in the absence of one, similar rebound-effect substitutes such as a sofa can be used~\cite{fang_vr_2023}. This allows for practicing the rhythm and procedure, with the intention of transitioning to a more realistic dummy when available. If no suitable object is available, a body visualization can be used in space, with both hands performing compressions on the digital representation in mid-air.

\paragraph{Backpack Sprayer [NF]}
The handle used to pressurize the tank is controlled by Nonlinear interactions, while the spray nozzle is controlled by Free interactions. For example, one hand can press the lever of a nail clipper to simulate pressurizing the tank (alternatively, this can be done by pressing the thumb against a bent index finger in mid-air, with the bending degree representing the pressure). The other hand can hold an object, such as a sunflower, and move it freely in space, with virtual liquid spraying from the center of the flower (or making an open-hand gesture in mid-air to represent the nozzle, with liquid spraying from the center of the palm). The spray speed depends on the tank's pressure, and when the pressure is insufficient, the other hand can continue to pressurize the tank.

\paragraph{Plant Watering [FB]}
As shown in Figures~\ref{fig:teaser-full}~(E1--3), the movement of the flowerpot is controlled by Free interactions, while watering is controlled by Binary interactions. For example, one hand can move and place an empty flowerpot anywhere in space (alternatively, in mid-air, the hand can form a hollow shape as if holding a small pot), and this physical pot will contain a virtual plant. The other hand can perform a pouring gesture above the plant; when this gesture is made, one unit (e.g., a small cup) of water is evenly distributed over the plant. When the gesture is not made, no water is poured (alternatively, pressing the button on the back of a pen can control the watering).

\paragraph{Breakwater and Tide [FL]}
The structure of the breakwater is controlled by Free interactions, while the size of the waves is controlled by Linear interactions. For example, one hand can manipulate tissue boxes, stacking them to different heights or shapes to represent and symbolize the breakwater's height (alternatively, virtual blocks can be grasped and arranged in mid-air). The other hand can slide along the edge of a table to control the wave intensity (alternatively, in mid-air, a single finger can slide vertically, with the height representing the wave intensity). This allows for observing how different breakwater structures and shapes respond to varying wave strengths.

\paragraph{Vacuum Cleaner [FR]}
The suction power of the vacuum cleaner is controlled by Rotational interactions, while the movement of the wand is controlled by Free interactions. For example, one hand can rotate a bottle cap, with the rotation angle representing the vacuum cleaner's speed/power (alternatively, a rotational gesture in mid-air can be used). The other hand, with a clenched fist and extended arm, represents the vacuum cleaner's wand and can move freely in space for cleaning (or a water bottle can be held, with a virtual wand attached to its neck, allowing the bottle's movement to control the wand for cleaning).

\paragraph{Parts Gluing [FN]}
The squeezing of the glue tube is controlled by Nonlinear interactions, while the movement of the objects being glued is controlled by Free interactions. For example, one hand can hold an empty small bottle (such as an empty eye drop bottle), and the squeezing degree controls the amount of glue applied (alternatively, this can be done with a mid-air gesture mimicking the squeezing of a glue bottle). The other hand can freely move the virtual parts to be glued into the desired position. The glue hand applies the glue, and the other hand holds the parts in place until the glue sets (alternatively, physical objects can replace virtual parts, such as using two tissue boxes).

\paragraph{Architectural Shadows [FF]}
As in the Shadow (\Cref{sec:shadow}), architects often consider the shadows cast by buildings in different seasons and times of the day to ensure proper light exposure. The position of the sun is controlled by Free interactions, and the positions of the buildings are also controlled by Free interactions. For example, one hand can manipulate coffee cup lids to represent buildings, moving them to different spatial positions (alternatively, in mid-air, the hand can mimic holding and moving a building). The other hand, making a fist, represents the sun and can move freely in space to affect the angle of sunlight, allowing observation of shadow changes (alternatively, one can hold a flashlight to represent the sun's light).

\paragraph{Marionette [FF-2]}
The two controls of a marionette are managed by two Free interactions. For example, one hand holds a forked branch and the other holds a straight branch, moving or swinging them in space to control the marionette's movements. If no suitable objects are available, mid-air gestures can be used instead---both hands are extended as if they are the controls, with strings connecting the hands to the marionette, allowing the hands to move and control the puppet in space. The marionette and its strings are virtual.

\endgroup

\end{document}